\documentclass[twocolumn]{aastex631}
\usepackage{bm}
\usepackage{amsmath}
\usepackage{mathtools,leftindex,tensor,mhchem}
\usepackage{comment}

\begin{document}

\title{Whistler-Alfv\'en turbulence in a non-neutral ultrarelativistic pair plasma}

\author[0000-0001-6252-5169]{Stanislav Boldyrev}
\affiliation{Department of Physics, University of Wisconsin at Madison, Madison, Wisconsin 53706, USA}
\affiliation{Center for Space Plasma Physics, Space Science Institute, Boulder, Colorado 80301, USA}

\author[0000-0001-5987-2856]{Mikhail Medvedev}
\affiliation{Department of Physics and Astronomy, University of Kansas, Lawrence, KS 66045, USA}
\affiliation{Laboratory for Nuclear Science, Massachusetts Institute of Technology, Cambridge, MA 02139, USA}



\begin{abstract}
The large-scale dynamics of most conventional space and astrophysical plasmas are predominantly governed by Alfv\'en modes, which are low-frequency magnetohydrodynamic modes existing in magnetized media. At scales smaller than the ion gyroscale or frequencies exceeding the ion cyclotron frequency, {the Alfv\'en modes transform into} kinetic-Alfv\'en or whistler modes that significantly contribute to plasma dynamics. However, this scenario reverses in non-neutral pair plasmas, such as those found in the magnetospheres of pulsars and magnetars, around rotating black holes and in their relativistic jets, as well as in certain laboratory plasmas. In these systems, the large-scale dynamics is governed by hybrid whistler-Alfv\'en modes, which transform into pure Alfvén modes at smaller scales. We derive the nonlinear equations that describe the dynamics of whistler-Alfv\'en modes in ultrarelativistic non-neutral magnetically dominated pair plasma and discuss the spectrum of turbulence governed by these equations.
\end{abstract}



\section{Introduction}
Pair plasmas are essential components of many high-energy astrophysical systems, including pulsar, magnetar, and black hole magnetospheres, jets from active galactic nuclei, and gamma-ray bursts \cite[e.g.,][]{arons1979,boettcher1997,sikora2000,meszaros2002,spitkovsky2006,petri2009,   philippov2014,philippov2015b,philippov2015a,geng2016,kawakatu2016,brambilla2018,blandford2019,hirotani2021}. Studies of collective plasma phenomena in these systems inspire efforts to create pair plasmas in laboratory experiments. Such experiments often involve interactions between powerful lasers and matter \cite[e.g.,][]{chen2023}, or the conversion of high-intensity, ultrarelativistic proton beams into electron-positron pairs through hadronic or electromagnetic cascades \cite[e.g,][]{arrowsmith2024}. In certain astrophysical environments, as well as in laboratory experiments, the pair plasma is not ideally neutral; in these cases, the electron densities can locally exceed the densities of positrons. The presence of uncompensated electric charge leads to significant modifications of collective effects in these plasmas \cite[e.g.,][]{maero2024}.

In this work, we are interested in plasma turbulence, which plays a significant role in various astrophysical systems, including the interstellar medium, stellar winds, and the magnetospheres of planets and stars. At scales much larger than the typical plasma microscales, such as ion gyroradii and ion inertial scales, the dynamics are effectively described by magnetohydrodynamic (MHD) Alfvén modes. These modes are low-frequency, large-scale oscillations that propagate along the background magnetic field. The nonlinear interaction of strong Alfv\'en waves leads to a turbulent energy cascade, which facilitates the transfer of energy from large-scale flows to smaller kinetic scales. In conventional plasmas, at scales smaller than the ion gyroscales or at frequencies higher than the ion gyrofrequencies, Alfv\'en modes are no longer present. Rather, the turbulent cascade in these regimes is governed by kinetic-Alfv\'en and whistler modes. These kinetic-scale modes are believed to be responsible for the ultimate dissipation of energy and the heating of plasma.



Conventional natural or laboratory plasmas, are, however, largely neutral because any local excess of electric charge is quickly neutralized by freely moving electrons. However, in certain important astrophysical environments and laboratory experiments, as discussed above, pair plasmas can be significantly non-neutral. We will demonstrate that in non-neutral pair plasmas, the dynamics display a contrasting behavior: at larger scales, the dynamics are dominated by whistler waves, which transition into Alfvén waves at smaller scales \cite[e.g.,][]{gedalin2001,urpin2011,medvedev2023}. In this work, we study the resulting whistler-Alfv\'en turbulence in non-neutral pair plasma. We derive the nonlinear two-fluid equations that govern the large-scale dynamics of non-neutral ultrarelativistic pair plasma and then discuss the spectra of strong turbulence that arise from these equations. 

The applicability of our results to specific situations depends on the plasma parameters of the systems studied. These parameters can vary significantly across different astrophysical environments as well as in current and planned laboratory experiments, necessitating careful consideration in each particular case. The objective of this work is rather to provide a general analysis of low-frequency waves and turbulence in a non-neutral pair plasma. As a specific example, we briefly examine pulsar magnetospheres and argue that non-neutrality in these systems obeys an important physical constraint, which ensures that it has only a marginal effect on Alfv\'enic modes.


\section{Linear waves} \label{sec:linear}
We start our discussion of turbulence with the derivation of linear low-frequency modes, see also \cite[][]{godfrey1975,arons1986,gedalin2001,urpin2011,medvedev2023,vega2024}, and then extend our analysis to the nonlinear dynamics. Consider a non-neutral ultrarelativistic pair plasma characterized by the non-neutrality parameter $|\Delta n_0|/n_0 \ll 1$, where $ \Delta n_0=n_0^+-n_0^-$ is the difference between the background positron and electron plasma densities, and $2 n_0=n_0^+ +n_0^-$ is the total lepton density. We assume that there are no net background currents associated with either species. 

{We also note that astrophysical plasmas can move along magnetic field lines at relativistic speeds. In these situations, our analysis is applicable to the rest frame of the plasma, where, for simplicity, we assume that the particle distribution function is symmetric with respect to the magnetic field direction. The frequencies and wavenumbers of the obtained modes can then be boosted to other reference frames. In more complex cases of multicomponent plasmas, where different components can move with different velocities and the plasma may also contain background currents, a similar approach can still be applied, although the resulting equations become more complicated  \cite[e.g.,][]{alexandrov84,gedalin1998,gedalin2001}.}

We consider the case of strongly magnetized and magnetically dominated plasma, so that the cyclotron frequency is much larger than the electron plasma frequency and  the frequency of the considered wave modes. 
In what follows, we denote $\omega_{pe}=\sqrt{4\pi n_0 e^2/m_e}$ the non-relativistic electron plasma frequency, $\Omega_e=eB_0/m_ec$ the non-relativistic electron cyclotron frequency, and $d_e=c/\omega_{pe}$ the non-relativistic electron inertial scale. We assume that the electron gyroscale is negligibly small,  $k_\perp^2\rho_{e}^2\ll 1$.  

We conventionally choose the coordinate frame such that the wave vector is given by ${\bf k}=(k_{\perp},0,k_{z})$, where $z$ is the direction of the background magnetic field. Under these assumptions, the plasma dielectric tensor is given by \cite[e.g.,][]{gedalin1998,gedalin2001,vega2024}:
\begin{eqnarray}
\varepsilon_{lm}=\left(\begin{array}{c c c}
1{~~} & -ig{~~} & 0\\
ig{~~} & 1{~~} & 0\\
0{~~} & 0{~~} & P
\end{array}\right),
\end{eqnarray}
where the function
\begin{eqnarray}
g(\omega)=\frac{\omega^2_{pe}}{\omega |\Omega_e|}\frac{\Delta n_0}{n_0}
\end{eqnarray}
is related to the breakdown of charge neutrality, and the function $P(\omega, {\bm k})$ depends on the particle distribution function; we will discuss it later. 

The dispersion relations and polarizations of the plasma waves are found from the wave equation:
\begin{eqnarray}
\left(k^{2}\delta_{lm}-k_{l}k_{m}-\frac{\omega^{2}}{c^{2}} \varepsilon_{lm}\right)E_{m}=0,
\end{eqnarray}
which in the matrix form reads
\begin{eqnarray}
\label{matrix}
{\left(\begin{array}{ccc}
k_{z}^{2}-\frac{\omega^{2}}{c^{2}}\quad\quad & i\frac{\omega^2}{c^2}g\quad & -k_{z}k_{\perp}\\
-i\frac{\omega^2}{c^2}g \quad & k^{2}-\frac{\omega^{2}}{c^{2}} \quad & 0 \\
-k_{z}k_{\perp}\quad & 0\quad & k_{\perp}^{2}-\frac{\omega^{2}}{c^{2}}P
\end{array}\right)}\left(\begin{array}{c}
E_{x}\\
E_{y}\\
E_{z}
\end{array}\right)=0.
\end{eqnarray}
Equating the determinant of the matrix to zero, we obtain:
\begin{multline}
\label{disp}
\left(k^{2}-\frac{\omega^{2}}{c^{2}}\right)\left[k_{\perp}^{2}-\frac{\omega^{2}}{c^{2}}P+k_z^2 P\right] \\
+\frac{\omega_{pe}^4}{c^2\Omega_e^2}\left(\frac{\Delta n_0}{n_0}\right)^2\left[k_\perp^2-\frac{\omega^2}{c^2}P\right]=0.\,\,\,\quad
\end{multline}

In order to proceed, we need to specify the function~$P$. 
A significant simplification occurs when considering an important case of  one-dimensional particle velocity distribution function, which assumes that the velocities of particles are nonzero only along the the background magnetic field. In this case, we get \cite[e.g.,][]{gedalin1998,vega2024}:
\begin{eqnarray}
\label{P}
P(\omega, {\bm k})=1-\frac{2\omega_{pe}^2}{\omega^2}\,W\left(\omega, k_z\right),
\end{eqnarray}
where the $W$ function is given by the standard integral:
\begin{eqnarray}
W=-\frac{\omega^2}{k_z}\int\limits_{-c}^{c}\frac{1}{\omega-k_zv_z +i\nu}\,\frac{d{\tilde f}}{dv_z}\,dv_z.
\label{W}
\end{eqnarray}
Here, $\nu\to +0$ is needed to take into account collisionless Landau damping. 

The one-dimensional distribution function is typically expressed through the variable $u_z=v_z/\sqrt{1-v_z^2/c^2}$, so that its normalization takes a simple form:
\begin{eqnarray}
\int\limits_{-\infty}^{\infty}{\tilde f}(u_z)du_z=1.    
\end{eqnarray}
For simplicity, we consider the one-dimensional Maxwell-J{\"u}ttner distribution,
\begin{eqnarray}
{\tilde f}(u_z)=\frac{1}{2cK_1(1/\vartheta)}\exp{\left(-\gamma/\vartheta\right)},    
\end{eqnarray}
where $K_1$ is the modified Bessel function of the second kind, $\gamma=1/\sqrt{1-v_z^2/c^2}=\sqrt{1+u_z^2/c^2}$ is the relativistic gamma-factor, and $\vartheta=k_BT/(m_ec^2)$ is the temperature parameter. We will consider the ultrarelativistic limit, $\vartheta\gg 1$, in which case $K_1(1/\vartheta)\approx \vartheta$.

The Landau damping is strong when $1-\omega^2/(k_z^2c^2)\sim 1/\vartheta^2$, which is the condition when the phase velocity of the wave is comparable to the thermal velocities of the particles.  In this case, the function $P$ has a large imaginary part, comparable to its real part. In contrast, its imaginary part is relatively small and can, therefore, be neglected in the following two limiting cases:
\begin{eqnarray}
\label{case1}
\left|1-\frac{\omega^2}{k_z^2 c^2}\right| \gg {1}/{\vartheta^2}\quad \mbox{(case I)},    
\end{eqnarray}
and 
\begin{eqnarray}
\label{case2}
\left| 1-\frac{\omega^2}{k_z^2 c^2}\right| \ll {1}/{\vartheta^2}\quad \mbox{(case II)}.    
\end{eqnarray}

As discussed in \cite[e.g.,][]{godfrey1975,gedalin1998,vega2024}, in case I, the $P$ function takes the form
\begin{eqnarray}
P\approx 1-\frac{2\omega_{pe}^2}{\vartheta\left(\omega^2-k_z^2c^2\right)},    
\end{eqnarray}
while in case II, we have
\begin{eqnarray}
P\approx 1-\frac{4\vartheta \omega_{pe}^2}{k_z^2c^2}.    
\end{eqnarray}
Obviously, in the case $\omega^2\geq k_z^2 c^2$, the imaginary part is absent, since the phase velocity of such waves is greater than the speed of light, and they are not affected by Landau damping.

In the presence of a strong guide field, plasma fluctuations associated with turbulence are typically anisotropic,~$k_\perp\gg k_z$. We are therefore interested in the low-frequency modes, $\omega^2<k_z^2c^2\ll \omega_{pe}^2/\vartheta$. Since we also consider a magnetically dominated plasma,~$\vartheta \omega_{pe}^2/\Omega_e^2\ll 1$, we always have $(\omega^2/c^2)|P|\gg k_\perp^2$.  Under these conditions, the dispersion relation (\ref{disp}) can be simplified:
\begin{eqnarray}
\label{disp_simp}   
k^2\left( k_z^2-\frac{\omega^2}{c^2}\right)=\frac{\omega^2}{c^2}\frac{\omega_{pe}^4}{c^2\Omega_e^2}\left(\frac{\Delta n_0}{n_0}\right)^2-\frac{k_\perp^2k^2}{P}.
\end{eqnarray}
The effects associated with the breakdown of plasma neutrality are presented by the first term in the right-hand side of this expression. The imaginary part and the associated effects of Landau damping are contained in the second term in the right-hand side.  

One can directly verify that in both case~I and case~II, the second term is much smaller than the first one if the wave number is restricted to the interval
\begin{eqnarray}
\label{w_limit}
k^2d_e^2 \ll \sqrt{\vartheta}\left(\frac{\omega_{pe}}{\Omega_e}\right)\left|\frac{\Delta n_0}{n_0}\right|.    
\end{eqnarray}
In this case, the Landau damping can be neglected, and the low-frequency mode has the dispersion relation analogous to that of the non-relativistic whistler or kinetic-Alfv\'en modes \cite[e.g.,][]{shukla2009,varma2011,chen_boldyrev2017,boldyrev2021},
\begin{equation}
\label{whistler}
\omega^2=\frac{k_z^2 c^2 k^2 d_e^2}{k^2d_e^2+ \left(\frac{\omega_{pe}}{\Omega_e}\right)^2\left(\frac{\Delta n_0}{n_0}\right)^2}.    
\end{equation}
Here, $d_e=c/\omega_{pe}$ is the non-relativistic electron inertial scale. 

Conversely, in the limit of large wave numbers,
\begin{eqnarray}
\label{w_limit2}
k^2d_e^2\gg \sqrt{\vartheta}\left(\frac{\omega_{pe}}{\Omega_e}\right)\left|\frac{\Delta n_0}{n_0}\right|,    
\end{eqnarray}
which formally corresponds to case~II, the last term in Eq.~(\ref{disp_simp}) becomes dominant, and mode (\ref{whistler}) has the dispersion relation consistent with the Alfv\'en mode:
\begin{eqnarray}
\label{alfven}
\omega^2=k_z^2c^2\left(1-\frac{k_\perp^2 c^2}{4\omega_{pe}^2\vartheta} \right).    
\end{eqnarray}

The Alfv\'en dispersion relation~(\ref{alfven}) holds while $k^2d_e^2\ll 4/\vartheta$, which follows from inequality~(\ref{case2}).  For larger wave numbers, the Alfv\'en mode is strongly dissipated by the collisionless Landau damping.

As seen from Eq.~(\ref{whistler}), the transition scale from the whistler-like behavior to Alfv\'en-like behavior is given by
\begin{eqnarray}
d_*=d_e\left(\Omega_e/{\omega_{pe}}\right)\left|n_0/{\Delta n_0}\right|.
\label{d*}
\end{eqnarray}
For this reason, we refer to this scale as the {\it whistler scale}. 
Additionally, as follows from Eqs.~(\ref{w_limit}) and~(\ref{w_limit2}), one can define another characteristic scale that corresponds to the transition between different sub-dominant asymptotics:
\begin{eqnarray}
d_{**}=d_e\left(\frac{\Omega_e}{\omega_{pe}}\right)^{1/2}\left|\frac{n_0}{\Delta n_0} \right|^{1/2}\vartheta^{-1/4}=\frac{\sqrt{d_e d_*}}{\vartheta^{1/4}}.~~
\end{eqnarray}
Since this scale is roughly the geometric mean of the inertial and whistler scales, so that both effects contribute, we refer to this scale as the {\it hybrid scale}.

\begin{figure}[h!]
\centering
\includegraphics[width=1.0\columnwidth]{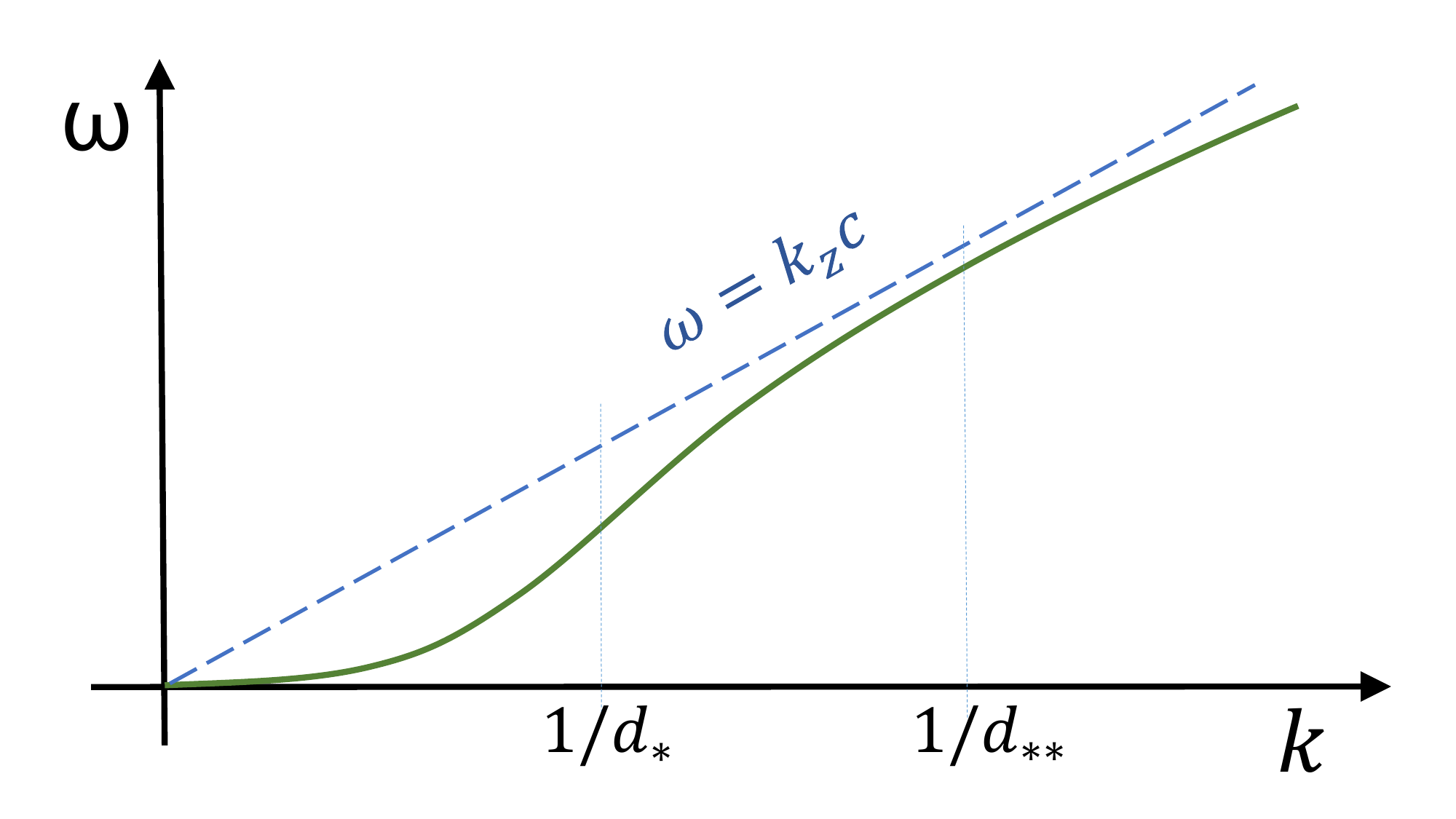}
\caption{Sketch of the dispersion relation of the mode described by Eqs.~(\ref{whistler}) and~(\ref{alfven}) for a given propagation angle. 
\label{dispersion}}
\end{figure}

We can then summarize the asymptotic behavior of the obtained modes in different scale intervals. At very large scales, $kd_*\ll 1$, the non-neutrality effects are significant, and we have the whistler-type mode:
\begin{eqnarray}
\label{whistler_large}
\omega^2\approx \Omega_e^2 k_z^2k^2d_e^4\left(\frac{n_0}{\Delta n_0}\right)^2\left(1-k^2d_*^2 \right).    
\end{eqnarray}
In the intermediate range of scales, $kd_*\gg 1$ but $kd_{**}\ll 1$, we have an Alfv\'en-type mode whose subdominant asymptotic (the dispersion correction) is defined by the non-neutrality effects. We will call this mode the hybrid whistler-Alfv\'en mode:
\begin{eqnarray}
\label{alfven-whistler}
\omega^2\approx k_z^2 c^2\left(1-\frac{1}{k^2d_*^2} \right).    
\end{eqnarray}
{The deviation of the dispersion relation from the linear form may be relevant, for instance, for the description of the parallel electric field as well as some types of nonlinear wave interactions.}
Finally, at small scales, $kd_{**}\gg 1$, we have a genuine Alfv\'en mode unmodified by the non-neutrality effects:
\begin{eqnarray}
\label{alfven_true}
\omega^2\approx k_z^2 c^2\left( 1-\frac{k_\perp^2d_e^2}{4\vartheta}\right).    
\end{eqnarray}
 The dispersion relation of the mode described by Eqs.~(\ref{whistler}) and~(\ref{alfven}) is sketched in Figure~\ref{dispersion}. 

The considered characteristic scales are much larger than the plasma inertial scale, $d_*\gg d_{**}\gg d_e$, but they may be much smaller than the typical outer scales characterizing astrophysical or laboratory system.  In such situations, in a non-neutral pair plasma, turbulence is whistler-like at large scales, but it becomes Alfv\'en-like at small scales.

\section{Nonlinear dynamics}
We now derive the equations governing nonlinear dynamics of the whistler-Alfv\'en mode. We consider plasma fluctuations that are anisotropic with respect to the strong background magnetic field, $k_z\ll k_\perp$. The nonlinear fluid-like equations can be derived using the procedure developed for the non-relativistic cases \cite[e.g.,][]{chen_boldyrev2017,loureiro2018,boldyrev2021,milanese2020} and adapted to the ultra-relativistic plasma in \cite[][]{vega2022b}. Here, we present the main steps of the derivation and discuss the difference introduced by the nonzero net charge of the plasma. 

Iterating the momentum equations for the electrons and positrons, one can derive the field-perpendicular velocities of the particles. The leading contribution comes from the E-cross-B drift, while the next order contribution in the small parameter $\omega\gamma_\alpha/\Omega_\alpha$, comes from the  polarization drift, 
\begin{equation}
{\bm v}_{\alpha, \perp} = {\bm v}_E +\frac{w_\alpha m_e c}{q_\alpha B^2}\,{\bm B}\times \frac{d_E}{dt}{\bm v}_E.
\label{vperp}
\end{equation}
Here, ${\bm v}_{E}=c({\bm E}\times{\bm B})/B^2$ is the E-cross-B velocity, $d_E/dt\equiv \partial/\partial t+{\bm v}_E\cdot{\bm \nabla_\perp}$ is the convective time derivative, $\alpha=\{e^+, e^-\}$ denotes the plasma species, and $w_\alpha n_\alpha m_ec^2$ is the enthalpy of each species. 

Note that we did not include in this equation the diamagnetic drift proportional to the perpendicular-pressure gradient, $\hat z\times \bm{\nabla}_\perp p_{\perp,\alpha}$, for two reasons. First, the diamagnetic drift does not lead to particle transport and it should not contribute to the continuity equation that we are going to use below. Second, it is reasonable to believe that the relevant particle distributions in a collisionless ultrarelativistic plasma correspond to a strongly anisotropic pressure tensor, with the field-perpendicular pressure being negligibly small as compared to the field-parallel one \cite[e.g.,][]{arons1986,gedalin1998,vega2024}. 

The field-parallel component of the velocity field is related to the field-parallel electric current,
$n_\alpha v_{\alpha, \|}={J_{\alpha, \|}}/{q_\alpha}$.
One then substitutes these perpendicular and parallel velocities into the continuity equations for the electrons and positrons, 
\begin{eqnarray}
\frac{\partial n_\alpha}{\partial t}+\bm{\nabla}_\perp\cdot \left(n_\alpha \bm{v}_{\alpha, \perp} \right)+\nabla_\|(n_\alpha v_\|)=0,
\end{eqnarray}
and obtains to the leading order in the small fluctuations:
\begin{multline}
\frac{\partial}{\partial t}\left(\frac{\delta n_\alpha}{n_{\alpha, 0}}-\frac{\delta B_z}{B_0}-\frac{w_{\alpha, 0}m_e c^2}{ q_\alpha B_0^2}\nabla_\perp^2\phi\right)\\
+\frac{1}{B_0}\left({\hat z}\times \bm{\nabla} \phi \right)\cdot \bm{\nabla} \left(\frac{\delta n_\alpha}{n_{\alpha, 0}}-\frac{\delta B_z}{B_0}-\frac{w_{\alpha, 0} m_ec^2}{q_\alpha B_0^2}\nabla_\perp^2\phi\right)\\
+\frac{1}{n_{\alpha, 0} q_\alpha}\nabla_\|J_{\alpha, \|}=0.
\label{eq:A}
\end{multline}
In this equation, $n_{\alpha, 0}$ denotes the background densities of each species, while $\delta n_\alpha=n_\alpha -n_{\alpha, 0}$ denotes the small density perturbations. 

Here, $\phi$ is the scalar electric potential, and ${\bm A}$ is the vector potential.
The field-parallel gradient in this equation is given by $
\nabla_\|=\partial/\partial z-\frac{1}{B_0}({\hat z}\times \bm{\nabla}_\perp A_z)\cdot \bm{\nabla}_\perp $, where the perpendicular magnetic field is expresses through the z-component of the vector potential, $\delta {\bm B}_\perp=-{\hat z}\times \bm{\nabla}_\perp A_z$.

Multiplying each of Equations (\ref{eq:A}) by $q_\alpha n_{\alpha, 0}$ and summing over the particle species, we obtain the charge conservation law:
\begin{multline}
\frac{\partial}{\partial t}\left({\rho}-\Delta \rho_0\frac{\delta B_z}{B_0}-\frac{2w_{0}n_0 m_e c^2}{ B_0^2}\nabla_\perp^2\phi\right) \\
+\frac{1}{B_0}\left({\hat z}\times \bm{\nabla} \phi \right)\cdot \bm{\nabla} \left({\rho}-\Delta \rho_0\frac{\delta B_z}{B_0}-\frac{2w_{0} n_0 m_e c^2}{  B_0^2}\nabla_\perp^2\phi\right) \\
+\nabla_\|J_{\|}=0,
\label{eq:A2}
\end{multline}
where $\rho=e^+\delta n^++e^-\delta n^-$ is the non-homogeneous (fluctuating) part of electric charge density, $J_\|=J^+_{\|}+J^-_{\|}$ is the parallel current, $w_0=\left(w^+_{0}+w^-_{0}\right)/2$ is the mean unperturbed enthalpy per particle, and $\Delta \rho_0=e^+n^+_{0}+e^-n^-_0=e^+\Delta n_0$ is the uniform background charge density of the plasma. We may assume that in the ultrarelativistic case, the unperturbed enthalpy is the same for both species, $w^+_{0}=w^-_{0}= w_0$. 

In the case of strong magnetization, $\sigma\gg 1$, however, we can neglect the $w_0$~containing terms altogether in Eq.~(\ref{eq:A2}) as they are small in  comparison with the charge density terms, $\rho=-\nabla^2\phi/4\pi$. We then get:
\begin{multline}
\frac{\partial}{\partial t}\left({\rho}-\Delta \rho_0\frac{\delta B_z}{B_0}\right) \\
+\frac{1}{B_0}\left({\hat z}\times \bm{\nabla} \phi \right)\cdot \bm{\nabla} \left({\rho}-\Delta \rho_0\frac{\delta B_z}{B_0}\right)
+\nabla_\|J_{\|}=0.
\label{eq:A3}
\end{multline}

In order to express $\delta B_z$, we consider the force balance in perpendicular momentum equation. Neglecting the small perpendicular pressure fluctuations, we may write:
\begin{eqnarray}
\Delta \rho_0{\bm E}=-\frac{1}{c}{\bm J}\times {\bm B}=-\frac{1}{4\pi}\left[{\bm \nabla}\times {\bm B}\right]\times {\bm B}.   
\end{eqnarray}
Recalling that $k_\perp \gg k_z$, and assuming that $\delta B_z\sim \delta B_\perp$ (this condition can be verified a posteriori), we obtain
\begin{eqnarray}
{\bm E}_\perp \approx \frac{B_0}{4\pi \Delta \rho_0}{\bm \nabla}_\perp \delta B_z.    
\end{eqnarray}
This equation allows one to express $\delta B_z$ in therms of the electric potential, ${\bm E}_{\perp}=-{\bm \nabla}_\perp \phi  $:
\begin{eqnarray}
\Delta \rho_0 \frac{\delta B_z}{B_0}=-\frac{4\pi e^2(\Delta n_0)^2}{B_0^2}\phi.    
\end{eqnarray}
Substituting this expression into Eq.~(\ref{eq:A3}), and expressing the charge density through the electric potential, we obtain:
\begin{multline}
 \frac{\partial}{\partial t}\left(-\frac{1}{4\pi}\nabla^2\phi+\frac{4\pi e^2(\Delta n_0)^2}{B_0^2}\phi\right) \\
+\frac{1}{B_0}\left({\hat z}\times \bm{\nabla} \phi \right)\cdot \bm{\nabla} \left(-\frac{1}{4\pi}\nabla^2\phi\right)
+\nabla_\|J_{\|}=0. 
\label{eq:A4}
\end{multline}

In order to close this equation, we need to know the electric current, which can be obtained from the field-parallel momentum equation:
\begin{eqnarray}
\frac{\partial {v}_{\alpha, \|}}{\partial t}+\left({\bm v}_E\cdot \bm{\nabla}_\perp \right){\bm v}_{\alpha, \|}= \frac{-{\nabla}_\| p_{\alpha}}{w_\alpha m_\alpha n_\alpha}+
\frac{q_\alpha}{w_\alpha m_\alpha}{E}_\|.\quad\quad
\label{eq:B}
\end{eqnarray}
Multiplying this equation by $n_{\alpha, 0}q_\alpha$ and summing over the particle species, we obtain:
\begin{eqnarray}
\frac{\partial {J}_\|}{\partial t}+\left({\bm v}_E\cdot \bm{\nabla}_\perp \right){J}_{\|}= \frac{- |e|}{w_0 m_e}{\nabla}_\| \delta p +
\frac{2 n_0 e^2}{w_0 m_e}E_\|,\quad
\label{eq:B2}
\end{eqnarray}
where $\delta p=\delta p^+-\delta p^-$ denotes the pressure imbalance. Assuming the adiabatic equation of state, the pressure variation can be expressed through the density variation, which, in tern, allows one to express the pressure imbalance through the electric potential:
\begin{eqnarray}
\delta p=w_0 m_e {v_s^2}\left(\delta n^+-\delta n^- \right)= -\frac{w_0 m_e {v_s^2}}{4\pi|e|}\nabla_\perp^2\phi, \quad\quad  
\end{eqnarray}
where $v_s$ is the acoustic speed in the ultrarelativistic pair plasma. 

Finally, one can express the electric current as $J_\|=-(c/4\pi)\nabla_\perp^2 A_z$, the parallel electric field as $E_\|=-\nabla_\|\phi-(1/c)\partial A_z/\partial t$, and substitute these expressions into Eq.~(\ref{eq:B2}). We then get a closed systems of two equations for the vector and scalar potentials.  In the resulting equations, we normalize the electric scalar potential and the $z$-component of the magnetic vector potential as ${\tilde \phi}={\phi} c/B_0$ and ${\tilde A}_z={A}_z c/(B_0\sqrt{1+2/\sigma})$. The perpendicular components of the magnetic and velocity fluctuations are expressed through these potentials as $\delta {\tilde{\bm b}}_\perp=-{\hat {\bm z}}\times \bm{\nabla} {\tilde A}_z$ and $\delta\tilde{{\bm v}}_\perp={\hat{\bm z}}\times \bm{\nabla} {\tilde \phi}$, and they both have the dimensions of velocity. Below we will use only these variables and omit the overtilde sign. 

Equations (\ref{eq:A4}) and (\ref{eq:B2}) then turn into a closed systems of equations describing low-frequency modes in a strongly magnetized, ultrarelativistic, non-neutral pair plasma:
\begin{multline}
\frac{\partial}{\partial t}\left( \Delta_n^2\phi -d^2_{rel}\nabla^2_\perp\phi \right)
-\left(\hat{{z}}\times \bm{\nabla}_{\perp} \phi \right)\cdot \bm{\nabla}_{\perp} d^2_{rel}\nabla^2_\perp\phi \\
=v_A\nabla_{\|} d^2_{rel}\nabla_{\perp}^{2} A_z, \quad \label{low_beta} 
\end{multline}
\begin{multline}
 \frac{\partial}{\partial t} \left( A_z -d_{rel}^2\nabla^2_\perp A_z\right) - \left(\hat{{z}}\times \bm{\nabla}_\perp\phi\right)\cdot\bm{\nabla}_\perp d_{rel}^2\nabla^2_\perp {A_z} \\
 = - v_A \nabla_\| \left( {\phi} -\frac{v_s^2}{c^2}d_{rel}^2\nabla_\perp^2\phi \right).
\label{reduced} 
\end{multline}
This system is the main result of this section. It describes the nonlinear dynamics of the hybrid Whistler-Alfv\'en fluctuations in a non-neutral pair plasma in the presence of a strong background magnetic field. 

In these equations, 
\begin{eqnarray}
v_A=\frac{c}{\sqrt{1+2/\sigma}}\approx c
\end{eqnarray}
denotes the Alfv\'en speed in a relativistic pair plasma, 
\begin{eqnarray}
d^2_{rel}=\frac{w_0}{2}\frac{c^2}{\omega_{pe}^2}=\frac{w_0m_ec^2}{8\pi n_0e^2}    
\end{eqnarray}
is the relativistic inertial scale, and the magnetic-field-parallel gradient is given by
\begin{eqnarray}
\label{nabla_par}
\nabla_\|=\partial/\partial z-\frac{1}{v_A}({\hat z}\times \bm{\nabla}_\perp A_z)\cdot \bm{\nabla}_\perp .
\end{eqnarray}
The plasma non-neutrality is described by the parameter
\begin{eqnarray}
\Delta_n^2=\frac{w_0}{2}\left(\frac{\omega_{pe}}{\Omega_e}\right)^2\left(\frac{\Delta n_0}{n_0}\right)^2   =\frac{1}{2\sigma }\left(\frac{\Delta n_0}{n_0}\right)^2,   
\end{eqnarray}
where 
\begin{eqnarray}
    \sigma=\frac{B_0^2}{4\pi n_0 w_0 m_e c^2},
\end{eqnarray}
is the plasma magnetization, and $w_0n_0 m_ec^2$ is the enthalpy of each species.  For the one-dimensional Maxwell-J\"uttner particle distribution considered here, the enthalpy is given by $w_0=K_2(1/\vartheta)/K_1(1/\vartheta)$. In the ultrarelativistic case, $\vartheta\gg 1$, this expression reduces to $w_0=2\vartheta$. When the plasma in neutral, $\Delta_n=0$, Equations (\ref{low_beta}) and (\ref{reduced}) describe Alfv\'enic dynamics and coincide with the equations previously studied in \cite[][]{vega2022b,vega2024}.

From Eqs.~(\ref{matrix}) and (\ref{reduced}), the relations between the magnitudes of the electric and magnetic fluctuations at large scales, $k_\perp^2 d_{rel}^2\ll \Delta_n^2$, are found as:
\begin{eqnarray}
E_\perp \sim \frac{k_\perp d_{rel}}{\Delta_n}  B_\perp.    
\end{eqnarray}
These electric fluctuations are much smaller than the magnetic fluctuations. Therefore, the large-scale whistler-type waves are mostly magnetic. In contrast, the Alfv\'en waves existing at $k_\perp^2 d_{rel}^2\gg \Delta_n^2$, are characterized by an equipartition between the magnetic and electric fluctuations, $E_\perp\sim  B_\perp$. 

\section{Whistler-Alfv\'en turbulence}
We now analyze low-frequency turbulence in a non-neutral pair plasma based on the system of equations~(\ref{low_beta}) and (\ref{reduced}). At hydrodynamic scales, $k_\perp d_{rel}\ll 1$, the system conserves the energy
\begin{eqnarray}
\label{Egeneral}
{\cal E}=\int\left\{\phi^2\Delta_n^2 +d_{rel}^2(\nabla_\perp \phi)^2 +d_{rel}^2(\nabla_\perp A_z)^2 \right\}d^3x,\quad\quad  
\end{eqnarray}
and helicity
\begin{eqnarray}
{\cal H}=    \int A_z\left(\phi \Delta_n^2-d_{rel}^2\nabla_\perp^2\phi \right)d^3x.
\end{eqnarray}

At scales $k_\perp d_e\gg \Delta_n$, the non-neutrality effects are not important, and the system describes the Alfv\'enic dynamics previously discussed in \cite[e.g.,][]{vega2022b,vega2024}. We, therefore, first concentrate on the  scales $k_\perp d_e\ll \Delta_n$, where the effects of non-neutrality dominate. 

In this limit, the system of equations transforms into the system mathematically analogous to that of reduced electron MHD (that describes the whistler modes) or kinetic-scale two-fluid Alfv\'enic equations (that describes the kinetic-Alfv\'en modes):
\begin{eqnarray}
\frac{\partial}{\partial t}\left( \Delta_n^2\phi \right) -v_A\nabla_{\|} d^2_{rel}\nabla_{\perp}^{2} A_z=0,  \label{eq1} \\
 \frac{\partial}{\partial t}  A_z + v_A \nabla_\| {\phi} =0.
\label{eq2} 
\end{eqnarray}
The nonlinearity in these equations is contained in the parallel gradients given by Eq.~(\ref{nabla_par}). 

First, we will discuss the standard scaling arguments allowing one to predict the energy spectrum of turbulence \cite[e.g.,][]{biskamp1999,cho2009} and then propose a refined model, specifically emphasizing the structures created by turbulence. {The dimensional arguments presented below are, therefore, also applicable to  conventional non-relativistic ion-electron plasmas.} 

The system given by Eqs.~(\ref{eq1}) and (\ref{eq2}) conserves the energy,
\begin{eqnarray}
{\cal E}=\int\left\{\phi^2\Delta_n^2 +d_{rel}^2(\nabla_\perp A_z)^2 \right\}d^3x ,  
\end{eqnarray}
and helicity,
\begin{eqnarray}
{\cal H}=  \Delta_n^2  \int \phi A_z \, d^3x.
\end{eqnarray}
In a turbulent state, the energy cascades toward small scales. The helicity would cascade to large scales and, therefore, it is not relevant for the energy spectrum at smaller scales that we discuss below. 

Turbulent eddies are in general anisotropic with respect to the local magnetic field, $l\gg \lambda$, where $l$~is their field-parallel dimension and $\lambda$ is the field-perpendicular dimension. We denote the corresponding typical fluctuations of the potentials as~$\phi_\lambda$ and~$A_{z,\lambda}$. The critical balance condition implies that the linear and nonlinear terms in the parallel gradient~(\ref{nabla_par}) are on the same order, 
\begin{eqnarray}
\label{crit}
v_A/l\sim A_{z,\lambda}/\lambda^2.
\end{eqnarray}
Balancing the other terms in the equations allows one to estimate~$\phi_\lambda\sim \left(d_{rel}/\Delta_n \right)A_{z,\lambda}/\lambda$. The time of nonlinear interaction evaluated from the nonlinear terms is then $\tau_\lambda \sim \lambda^2/\phi_\lambda$. Since the energy of fluctuations scales as ${\cal E}_\lambda \sim \Delta_n^2 \phi_\lambda^2$, we estimate the energy flux as
\begin{eqnarray}
\epsilon\sim {\cal E}_\lambda/\tau_\lambda\sim  \phi_\lambda^3\Delta_n^2/\lambda^2.    
\end{eqnarray}
The energy flux, $\epsilon$, should be independent of scale, which gives $\phi_\lambda\sim \epsilon^{1/3}\left(\lambda/\Delta_n\right)^{2/3}$, leading to the Fourier energy spectrum 
\begin{eqnarray}
\label{spec1}
{\cal E}_{k_\perp} dk_\perp \propto \epsilon^{2/3}\Delta_n^{2/3} k_{\perp}^{-7/3}dk_\perp.  
\end{eqnarray}
The anisotropy of turbulent eddies is then obtained from Eq.~(\ref{crit}): 
\begin{eqnarray}
\label{anis}
l\propto \lambda^{1/3}.  
\end{eqnarray}

Numerical simulations and solar wind observations, however, demonstrate that the spectrum of kinetic-Alfv\'en waves is steeper than that given by Eq.~(\ref{spec1}) \cite[e.g.,][]{leamon1998,bale_measurement_2005,sahraoui2009,salem2012,boldyrev12b,tenbarge2012,wan2015,told2015,servidio2015,cerri2018,groselj2018,arzamasskiy2019,zhou2023}. Two potential reasons have been investigated for the spectral steepening. First is the effects of Landau damping which is generally present in an electron-ion plasma at kinetic scales \cite[e.g.,][]{tenbarge2012}. Second is the intermittency of turbulence, which implies that the energy containing regions are not space filling but rather occupy a (generally fractal) volume with the dimension~$D<3$ \cite[e.g.,][]{boldyrev_p12,zhou2023}. 

Here, we analyze the second scenario, and argue that the spectrum of strong kinetic-Alfv\'en or whistler turbulence should be steeper than~$k_\perp^{-5/2}$. First, we  present phenomenological arguments indicating that the spectrum $k_\perp^{-7/3}$ is not realizable. For that we estimate the curvature of magnetic-field lines associated with a turbulent eddy predicted by Eq.~(\ref{anis}),
\begin{eqnarray}
K_\lambda\sim \lambda/l^2 \propto \lambda^{1/3}.    
\end{eqnarray}
We see that if the scaling~(\ref{anis}) were true, the curvature of a magnetic-field line would {\it decrease} as the scale of the eddies decreases. This means that  eddies of scale~$\lambda$ propagating along the magnetic field lines would be stronger distorted by the large-scale eddies than by the eddies of a comparable size. Therefore, the critical balance condition~(\ref{crit}) would not be satisfied. The eddy distortion time would be shorter and the energy spectrum would be steeper, which leads to a contradiction. 

We now argue that intermittency of turbulence can remove this contradiction. To demonstrate this, we follow the procedure of \cite[][]{boldyrev12b} and assume that that the fraction of the two-dimensional volume occupied by eddies in the field-perpendicular plane scales as~$p_\lambda\propto \lambda^{3-D}$, where~$1<D<3$. The energy is then estimated as ${\cal E}_\lambda \propto \phi_\lambda^2p_\lambda$, and the energy flux is
\begin{eqnarray}
{{\cal E}_\lambda}/{\tau_\lambda}\propto {\phi_\lambda^3p_\lambda}/{\lambda^2}.    
\end{eqnarray}
The energy flux is independent of scale if $\phi_\lambda \propto \lambda^{(D-1)/3}$, which leads to the energy spectrum
\begin{eqnarray}
\label{Efinal}
{\cal E}_{k_\perp}dk_\perp \propto k_\perp^{(D-10)/3}dk_\perp.     
\end{eqnarray}
The anisotropy of the eddies is then given by
\begin{eqnarray}
l\propto \lambda^{(4-D)/3},    
\end{eqnarray}
which implies that the magnetic field-line curvature associated with the eddy scales as
\begin{eqnarray}
K_\lambda \propto \lambda^{(2D-5)/3}.    
\end{eqnarray}

A self-consistent turbulent model should require that the curvature {\it increases} as the scale decreases, which leads to the realizability  condition that restricts the dimensionality of energy containing regions to~$1<D<5/2$. We, therefore, expect that the energy spectrum of strong kinetic-Alfv\'en or whistler turbulence should be bound between 
$ {\cal E}_{k_\perp}\propto k_\perp^{-5/2} $ and ${\cal E}_{k_\perp}\propto k_\perp^{-3}$, see Eq.~(\ref{Efinal}), which is consistent with the observational and numerical studies cited above. The corresponding anisotropy of turbulent fluctuations is then restricted between $l\propto \lambda^{1/2}$ and $l\propto \lambda$. 

At small scales, $k_\perp d_{rel}\gg \Delta_n$, the whistler-like modes transform into the Alfv\'en modes, and the system of equations~(\ref{low_beta}), (\ref{reduced}) transforms into the well-known equations of reduced MHD \cite[e.g.,][]{biskamp2003}:
\begin{eqnarray}
\frac{\partial}{\partial t}\left( \nabla_\perp^2\phi \right) +v_A\nabla_{\|} \nabla_{\perp}^{2} A_z=0,  \label{eq3} \\
 \frac{\partial}{\partial t}  A_z + v_A \nabla_\| {\phi} =0.
\label{eq4} 
\end{eqnarray}
The energy integral transforms into the expression describing the total (magnetic plus electric) energy,
\begin{eqnarray}
{\cal E}=d_{rel}^2\int\left\{\left(\nabla_\perp \phi\right)^2 +(\nabla_\perp A_z)^2 \right\}d^3x,   
\end{eqnarray}
while the helicity transforms into the integral of the so-called ``cross-helicity'':
\begin{eqnarray}
{\cal H}=    d_{rel}^2\int {\bm \nabla}_\perp\phi \cdot {\bm \nabla}_\perp A_z \, d^3x.
\end{eqnarray}

The spectrum of Alfv\'enic turbulence has been studied in both conventional and relativistic plasmas \cite[see, e.g.,][]{goldreich_toward_1995, boldyrev2006,mason2006,mason2012,boldyrev2009,perez_etal2012,tobias2013,mallet2015,chandran_intermittency_2015,chen2016,walker2018,kasper2021,chernoglazov2021,vega2022b}. In the presence of a strong guide magnetic field, the spectrum scales approximately as:
\begin{eqnarray}
{\cal E}_{k_\perp} dk_\perp\propto k_\perp^{-3/2}\,dk_\perp,  
\end{eqnarray}
which is significantly flatter than the spectrum of the whistler turbulence.

An interesting difference of the present case with the previous studies is that the helicity flux from large to small scales is absent in the whistler turbulent cascade existing at scales $k_\perp d_{rel}\ll \Delta_n$. Therefore, at the smaller Alfv\'enic scales, the resulting cross-helicity is negligibly small. This means that the Alfv\'enic turbulence appearing at scales $k_\perp d_{rel}\gg \Delta_n$ is balanced, that is, it has comparable fluxes of energy propagating along and against the background magnetic field. 

As mentioned in the introduction, non-neutral pair plasmas can exist in a variety of astrophysical and laboratory environments.  The analysis of the nonlinear turbulent properties of these systems is sensitive to the values of the characteristic scales $d_*$ and $d_{**}$. Below, we briefly discuss the parameters of non-neutral pair plasma in the context of pulsar and magnetar magnetospheres as an example.

\section{Pulsar and magnetar plasma}


Pulsars and magnetars are types of neutron stars that form from the remnants of core-collapse supernovae. These supernovae typically originate from massive stars with masses ranging from approximately 8 to 12 solar masses ($M \sim 8-12 M_\odot$). Neutron star masses generally fall within a narrow range of about $1.18 M_\odot$ to $2 M_\odot$, with an average mass around $M_* \sim 1.35 M_\odot$. In contrast to their masses, the spins of neutron stars vary widely, spanning over three orders of magnitude. Young pulsars and magnetars can have rotation periods measured in seconds, while older ``recycled'' pulsars can spin with millisecond periods. Neutron stars also possess extraordinarily strong magnetic fields. The magnetic field strength at the surface, denoted as $B_*$, typically ranges from $10^8$ to $10^{12}$ Gauss for ordinary pulsars. However, some neutron stars can exceed the Schwinger (quantum) field strength of approximately $B_{Q} = m_e^2c^2/e\hbar \sim 4 \times 10^{13}$ gauss and can have magnetic fields as strong as $B_* \sim 10^{15}$ gauss. The latter class of neutron stars is known as magnetars.

It is generally accepted that the magnetosphere of pulsars and magnetars can be described by a dipolar magnetic field up to the "light cylinder" (LC). The LC is defined as the region where the linear speed approaches the speed of light, specifically where $\Omega \, r_{\rm LC} \sim c$, with $\Omega$ representing the angular velocity of the pulsar's rotation. When a pulsar is modeled as a spinning, magnetized conducting sphere, a large electric field component appears parallel to the magnetic field. Consequently, a free electrically charged particle could attain unrealistically high energy levels unless this electric field is screened. \citet{GJ69} proposed that such screening occurs due to a plasma filling the magnetosphere, which causes it to corotate with the neutron star. 

To satisfy the force-free condition, the equation ${\bm E} + \left( \frac{{\bm v}}{c} \right) \times {\bm B} = 0$ must hold true. In the laboratory frame, the corotation electric field can be expressed as ${\bm E}_{\rm cor} = -\left( \frac{1}{c} \right) \left( {\bm \Omega} \times {\bm r} \right) \times {\bm B}$, where ${\bm r}$ is the position vector. 

From the Poisson equation, it can be inferred that to maintain this corotation, the magnetosphere must be locally filled with plasma that possesses the so-called "Goldreich-Julian" charge density.
\begin{equation}
    \rho_{\rm GJ}=\frac{1}{4\pi}{\bm \nabla}\cdot {\bm E}_{\rm cor}
    =-\frac{{\bm \Omega}\cdot{\bm B}}{2\pi c}
    +\frac{{\bm \Omega}\cdot\left({\bm r}\times({\bm \nabla}\times{\bm B})\right)}{4\pi c}.
\end{equation}
For a purely dipolar magnetic field, the second term vanishes because $\bm{\nabla} \times \bm{B} = 0$. It is important to note that this plasma is non-neutral and carries uncompensated electric charge. The boundary between the positively and negatively charged regions is determined by the surface where $\bm{\Omega} \cdot \bm{B} = 0$. The corresponding Goldreich-Julian (GJ) particle density is given by the following equation:
\begin{equation}
    n_{\rm GJ}=\rho_{\rm GJ}/e\approx\frac{\Omega B}{2\pi ce}\simeq0.07\frac{B}{P_*}\textrm{ cm}^{-3},
\end{equation}
where $B$ is the magnetic field strength in gauss and $P_*=2\pi/\Omega$ is the spin period in seconds. 

For example, for the Crab pulsar, which has the period of $P\simeq0.0335$~s and the magnetic field $B(r)\simeq4\times10^{12}(r/R_*)^{-3}$~gauss ($R_*$ being the neutron star radius), the plasma density is
\begin{equation}
n_{e}(r)\simeq8.5 \times10^{12}\ {\cal M} \left(\frac{r}{R_*}\right)^{-3}\textrm{ cm}^{-3}.
\end{equation}
Here we introduced a `multiplicity' factor ${\cal M}$, which accounts for the possibility that the plasma density can exceed the GJ limit (see discussion below). 

One should note that the GJ estimate is valid for the (inner) magnetosphere inside the light cylinder. Outside the LC, the magnetosphere cannot corotate with the star. It was shown (e.g., see review by \citealp{philippov-kramer22}) that the magnetosphere `opens up' and forms a `split monopole' configuration with a large $\phi$-component. The $r$- and $\phi$-components of the magnetic field outside the LC are $B_r=\pm B_*(r/R_*)^{-2}, \ B_\phi=-(\Omega r/c)B_r$. The hemispheres with the positive and negative magnetic flux are separated by a thin current sheet in the equatorial plane. The plasma density in the wind region, outside the light cylinder, scales as $n_e\propto r^{-2}$. The above description is most applicable for the aligned rotator when the magnetic and rotation axes coincide. The field and current sheer configurations become more complicated and dynamic for an inclined rotator and, especially, an orthogonal rotator.  

\citet{timokhin10} has shown that the magnetosphere is populated with plasma via an intrinsically unsteady discharge process. It involves the repeated cycle of particle acceleration, curvature radiation, electron-positron creation via $\gamma+B\to e^+ +e^-$, and subsequent evacuation of the ``spark gap'' region, which further creates the accelerating parallel electric field. This mechanism applies to most pulsars except the very young ones, possessing strong (megagauss) magnetic fields at the light cylinder. The available free energy comes from the rotational energy of the neutron star. Numerical simulations indicate that the above process leads to the production of the strong plasma wind along the open field lines, with typical Lorentz factors of the primaries $\gamma\sim10^6-10^7$ and secondaries $\gamma\sim10^1-10^3$, and multiplicities ${\cal M}=n/n_{\rm GJ}\sim10^3-10^5$ \citep{philippov-timokhin20,philippov-kramer22,sashaCh+24}. This wind carries along the (observed) Poynting flux, causing magnetic braking of the pulsar. We stress that the plasma remains electrically non-neutral, despite large multiplicity. The population of the inner magnetosphere (the dipolar, closed field line structure) is likely happening via cross-L-shell diffusion from the separatrix and the Y-point, possibly along with direct pair production by $\gamma$-$\gamma$ interaction. Since no wind appears here, the plasma is `calm' and has enough time to cool and relax. Hence, one expects the thermal Lorenz factor to be $\langle\gamma\rangle\sim1$ and lower multiplicity ${\cal M}\sim 1-100$.

Plasma in a magnetar magnetosphere is believed to have somewhat different parameters. In magnetars, the magnetosphere can twisted by surface shear (Hall) motions, so it is threaded by electric currents ${\bm j}=(c/4\pi){\bm \nabla}\times{\bm B}$. Thus, the magnetosphere should carry electron-positron plasma needed to maintain the current, $n=j/c e\simeq (10^{17}{\rm cm}^{-3})(B/10^{15}\textrm{ gauss})(r/10^{6}\textrm{ cm})$. For comparison, a typical magnetar spins with a period of 1 second and carries a surface field of at most $10^{15}$ gauss. This yields the GJ density $n_{\rm GJ}\sim 10^{14}\textrm{~cm}^{-3}$. The GJ current $j_{\rm GJ}\sim\rho_{\rm GJ}c$ is not enough to maintain the twist, so an electron-positron cascade is again required \citep{beloborodov-thompson07}. Effectively, the inner magnetar magnetospheric plasma is weakly non-neutral (in order to corotate) with the multiplicity of about ${\cal M}\sim10^3$ and a mean Lorentz factor $\langle\gamma\rangle\sim10^3$.

Finally, the neutron crust is a conducting solid with free electrons. The electron density varies between $n_e\sim 2.5\times\sim 10^{36}\textrm{~cm}^{-3}$ at the liquid core--solid crust interface and $n_e\sim 3\times 10^{31}\textrm{~cm}^{-3}$ at the neutron star surface \citep{pearson+11,potekhin+13}. At a typical temperature of a young neutron star of $T\sim10^8$~K, the conductivity varies from $\sigma_c\sim 10^{25}\textrm{~s}^{-1}$ at the core-crust interface and $\sigma_c\sim 7\times10^{19}\textrm{~s}^{-1}$ at the surface. In this system, the positive ions are motionless (except for glitches) and the magnetic field is advected by the degenerate electrons. The tension of the magnetic field lines at the core-crust interface generates whistler waves which accumulate in the crust and provide strong maxwell stress that can lead to glitches \citep{bransgrove+24}.

\section{Estimates of ${\MakeLowercase d}_*$ and ${\MakeLowercase d}_{**}$ for pulsars and magnetars}

In the inner magnetospheres of pulsars and magnetars, the charge non-neutrality is entirely associated with the GJ density $\Delta n_0=n_{\rm GJ}$, even though the total plasma density is larger by the multiplicity factor, $n_0={\cal M}n_{\rm GJ}$. Taking this into account, Eq. (\ref{d*}) becomes
\begin{eqnarray}
d_*=\frac{B}{4\pi e \left|n_{\rm GJ}\right|}=\frac{R_{\rm LC}}{2\pi}. 
\label{d*inner}
\end{eqnarray}
Note, this estimate is insensitive to the plasma multiplicity. 

In contrast, the inertial scale and the hybrid scale do depend on plasma parameters and location
\begin{eqnarray}
d_e&=&\frac{c}{\omega_{pe}}
=(0.18\textrm{ cm})\left(\frac{P_* r_6^3}{B_{12}{\cal M}}\right)^{1/2}, 
\\
d_{**}&=&\frac{\sqrt{d_e d_*}}{\vartheta^{1/4}}
=(3.9\times10^4\textrm{ cm})\left(\frac{P_*^{3} r_6^3}{B_{12}{\cal 
M}\vartheta}\right)^{1/4},
\label{d**inner}
\end{eqnarray}
where $B_{12}=B/(10^{12}\textrm{G})$, $r_{6}=B/(10^{6}\textrm{cm})$ (assuming $R_*=10$~km), $P_*$ is the star spin period in seconds. In the inner magnetosphere of pulsars, the plasma is relatively cold and not too dense, $\vartheta\sim\langle\gamma\rangle\sim1$ and ${\cal M}\sim 10$. In magnetars, in turn, $\vartheta\sim\langle\gamma\rangle\sim10^3$, ${\cal M}\sim 10^3$, and $B_{12}\sim1000$. Thus, $d_{**}$ in magnetars is smaller than in pulsars by nearly two orders of magnitude. We emphasize that in a neutron star inner magnetosphere $d_{**}$ is much smaller than the radius of the star, $d_{**}\ll R_*$, at the star surface. Since $d_{**}\propto r^{3/4}$, $d_{**}$ is always smaller than the current radius in the system, $d_{**}\ll r$.

The outer magnetosphere, $r\gg R_{LC}$, is filled with the pulsar wind. Here, the charge non-neutrality can be estimated as follows. 

First, we note that the plasma velocity is represented by the free motion along the magnetic field plus the corotation velocity, ${\bm v}=v_\| \hat b + {\bm \Omega}\times{\bm r}=v_\| {\hat b} + \Omega r \sin\theta \hat\phi$. From the ideal MHD condition ${\bm E}=-\left(\frac{{\bm v}}{c}\right)\times {\bm B}$, we obtain that the electric field is largely poloidal, ${\bm E}= E_\theta \hat \theta$ where
\begin{equation}
    E_\theta=\frac{\Omega r}{c}\sin\theta B_r.
\end{equation}

Second, asymptotically, at $r\sin\theta\gg R_{LC}$, the magnetic field is highly wound-up and dominated by the toroidal component, $B_\phi\gg B_r, B_\theta$. The wind is moving with the drift velocity $v_{E\times B}=c\left|{\bm E}\times{\bm B}\right|/B^2\approx c |E_\theta B_\phi|/B_\phi^2$. Since, the wind is relativistic $v_{E\times B}\approx c$, one has the relation $B_\phi=-E_\theta$.

Next, the charge non-neutrality is associated with the poloidal electric field, similarly to the GJ argument. Indeed, the non-neutral plasma density follows from  Poisson's equation, $n_\theta={\bm \nabla}\cdot{\bm E}_\theta/(4\pi e)$, to yield 
\begin{equation}
n_\theta=\frac{\Omega B_r\cos\theta}{2\pi c e}. 
\end{equation}
The total lepton density is as usual $n_0=n_\theta{\cal M}$ and $B_r=B_*(R_{\rm LC}/R_*)^{-3}(r/R_{\rm LC})^{-2}$.

Finally, noting that $B\approx B_\phi$, we obtain  
\begin{eqnarray}
d_*\approx\frac{B_\phi}{4\pi e n_\theta}
=\frac{1}{2}\,r\tan\theta, 
\label{d*outer}
\end{eqnarray}
which, again, is independent of the plasma multiplicity and the $B_r$ configuration. 
In contrast, 
\begin{eqnarray}
d_e&=&(1.4\times10^5\textrm{ cm})\frac{P_* r_9}{\left(B_{12}{\cal M}\cos\theta\right)^{1/2}}, 
\\
d_{**}&=&(1.4\times10^{12}\textrm{ cm})\frac{P_*^{1/2} r_9 (\tan\theta)^{1/2}}{\left(B_{12}{\cal M}\vartheta\cos\theta\right)^{1/4}},
\end{eqnarray}
where $r_{9}=B/(10^{9}\textrm{cm})$. We note that, as before, $d_{**}\ll r$.

We conclude that in the environments we considered, the whistler scale $d_* $ is nearly always comparable to the largest scale of the system, which leaves little room for dynamics dominated by whistler modes. At the same time, the hybrid scale $d_{**} $ is consistently very small, indicating that the effects of non-neutrality significantly influence the dispersive properties of the modes. Consequently, we conclude that at nearly all scales relevant to pulsar and magnetar magnetospheres, the dispersion of low-frequency Alfv\'en modes is affected by non-neutrality. As a result, the Alfv\'en modes are replaced by the hybrid-Alfv\'en modes, as described by Eqs.~(\ref{whistler}) and~(\ref{alfven-whistler}).



\section{Summary}
We examined the dynamics of large-scale, low-frequency fluctuations in a non-neutral ultra-relativistic pair plasma that is embedded in a strong background magnetic field. At scales larger than a certain characteristic scale (the whistler scale), denoted as $d_*$, the dynamics resemble those of whistlers waves, similar to those typically found in conventional plasmas. In the intermediate range of scales, between $d_*$ and $d_{**}$, the dynamics are characterized by what we called whistler-Alfv\'en modes, while at scales smaller than the hybrid scale $d_{**}$, these modes transition to the genuine Alfv\'en mode. 

We have derived a set of two-fluid equations that provide a unified description of nonlinear plasma dynamics across all three regimes. Using these equations, we analyzed the spectra of strong turbulence resulting from the energy cascade from large to small scales. Our findings may have implications for plasma conditions in the magnetospheres of pulsars and magnetars, around rotating black holes and in their relativistic jets, as well as in certain laboratory plasmas that contain a significant non-neutral pair plasma component.

{The applicability of our results to specific situations depends on the plasma parameters of the systems studied, which can vary significantly across different astrophysical environments and laboratory experiments. As an example, we examined the parameters of the magnetospheres of pulsars and magnetars, where the non-neutrality of the plasma is caused by the rotation of conducting magnetized plasmas. We demonstrated that in such cases, the characteristic ``whistler scale" $d_* $ is comparable to the size of the light cylinder, while the hybrid scale $ d_{**} $ is smaller than the radius of the star. It would be interesting to extend similar analysis to other physical environments, including the cases where the plasma or its components can move with relativistic velocities along the background magnetic fields and the plasma may also contain net currents. We leave these questions for future studies.}

\begin{acknowledgments}
We are grateful to the anonymous reviewer for useful comments. The work of SB was supported by the U.S. Department of Energy, Office of Science, Office of Fusion Energy Sciences under award number DE-SC0024362. The work of MM was supported by the National Science Foundation under Grant No. PHY-2409249. This research was also supported in part by grant NSF PHY-2309135 to the Kavli Institute for Theoretical Physics (KITP).
\end{acknowledgments}

\bibliography{references}{}

@BOOK{alexandrov84,
       author = {{Alexandrov}, A. F. and {Bogdankevich}, L. S. and {Rukhadze}, A. A.},
        title = "{Principles of Plasma Electrodynamics}",
    publisher = {Springer Berlin, Heidelberg},
         year = 1984,
          doi = {},
       adsurl = {}
}

@ARTICLE{medvedev2023,
       author = {{Medvedev}, Mikhail V.},
        title = "{Plasma modes in QED super-strong magnetic fields of magnetars and laser plasmas}",
      journal = {Physics of Plasmas},
         year = 2023,
        month = sep,
       volume = {30},
       number = {9},
          eid = {092112},
        pages = {092112},
          doi = {10.1063/5.0160628},
archivePrefix = {arXiv},
       eprint = {2309.07316},
 primaryClass = {physics.plasm-ph},
       adsurl = {https://ui.adsabs.harvard.edu/abs/2023PhPl...30i2112M}
}

@article{geng2016,
doi = {10.3847/0004-637X/825/2/107},
url = {https://dx.doi.org/10.3847/0004-637X/825/2/107},
year = {2016},
month = {jul},
publisher = {The American Astronomical Society},
volume = {825},
number = {2},
pages = {107},
author = {Geng, J. J. and Wu, X. F. and Huang, Y. F. and Li, L. and Dai, Z. G.},
title = {IMPRINTS OF ELECTRON–POSITRON WINDS ON THE MULTIWAVELENGTH AFTERGLOWS OF GAMMA-RAY BURSTS},
journal = {The Astrophysical Journal}
}

@article{blandford2019,
   author = "Blandford, Roger and Meier, David and Readhead, Anthony",
   title = "Relativistic Jets from Active Galactic Nuclei", 
   journal= "Annual Review of Astronomy and Astrophysics",
   year = "2019",
   volume = "57",
   number = "Volume 57, 2019",
   pages = "467-509",
   doi = "https://doi.org/10.1146/annurev-astro-081817-051948",
   url = "https://www.annualreviews.org/content/journals/10.1146/annurev-astro-081817-051948",
   publisher = "Annual Reviews",
   issn = "1545-4282",
   type = "Journal Article",
  }

@article{sikora2000,
doi = {10.1086/308756},
url = {https://dx.doi.org/10.1086/308756},
year = {2000},
month = {may},
publisher = {},
volume = {534},
number = {1},
pages = {109},
author = {Sikora, M. and Madejski, G.},
title = {On Pair Content and Variability of Subparsec Jets in
Quasars},
journal = {The Astrophysical Journal}
}

@ARTICLE{kawakatu2016,
       author = {{Kawakatu}, Nozomu and {Kino}, Motoki and {Takahara}, Fumio},
        title = "{Evidence for a significant mixture of electron/positron pairs in FRII jets constrained by cocoon dynamics}",
      journal = {\mnras},
         year = 2016,
        month = mar,
       volume = {457},
       number = {1},
        pages = {1124-1136},
          doi = {10.1093/mnras/stw010},
archivePrefix = {arXiv},
       eprint = {1601.00771},
 primaryClass = {astro-ph.HE},
       adsurl = {https://ui.adsabs.harvard.edu/abs/2016MNRAS.457.1124K}
}

@ARTICLE{boettcher1997,
       author = {{Boettcher}, M. and {Mause}, H. and {Schlickeiser}, R.},
        title = "{{\ensuremath{\gamma}}-ray emission and spectral evolution of pair plasmas in AGN jets. I. General theory and a prediction for the GeV - TeV emission from ultrarelativistic jets.}",
      journal = {\aap},
         year = 1997,
        month = aug,
       volume = {324},
        pages = {395-409},
          doi = {10.48550/arXiv.astro-ph/9604003},
archivePrefix = {arXiv},
       eprint = {astro-ph/9604003},
 primaryClass = {astro-ph},
       adsurl = {https://ui.adsabs.harvard.edu/abs/1997A&A...324..395B}
}

@article{brambilla2018,
doi = {10.3847/1538-4357/aab3e1},
url = {https://dx.doi.org/10.3847/1538-4357/aab3e1},
year = {2018},
month = {may},
publisher = {The American Astronomical Society},
volume = {858},
number = {2},
pages = {81},
author = {Brambilla, Gabriele and Kalapotharakos, Constantinos and Timokhin, Andrey N. and Harding, Alice K. and Kazanas, Demosthenes},
title = {Electron–Positron Pair Flow and Current Composition in the Pulsar Magnetosphere},
journal = {The Astrophysical Journal}
}

@article{philippov2014,
doi = {10.1088/2041-8205/785/2/L33},
url = {https://dx.doi.org/10.1088/2041-8205/785/2/L33},
year = {2014},
month = {apr},
publisher = {The American Astronomical Society},
volume = {785},
number = {2},
pages = {L33},
author = {Philippov, Alexander A. and Spitkovsky, Anatoly},
title = {AB INITIO PULSAR MAGNETOSPHERE: THREE-DIMENSIONAL PARTICLE-IN-CELL SIMULATIONS OF AXISYMMETRIC PULSARS},
journal = {The Astrophysical Journal Letters}
}

@article{spitkovsky2006,
doi = {10.1086/507518},
url = {https://dx.doi.org/10.1086/507518},
year = {2006},
month = {aug},
publisher = {},
volume = {648},
number = {1},
pages = {L51},
author = {Spitkovsky, Anatoly},
title = {Time-dependent Force-free Pulsar Magnetospheres: Axisymmetric and Oblique Rotators},
journal = {The Astrophysical Journal}
}

@ARTICLE{philippov2015a,
       author = {{Philippov}, Alexander A. and {Spitkovsky}, Anatoly and {Cerutti}, Benoit},
        title = "{Ab Initio Pulsar Magnetosphere: Three-dimensional Particle-in-cell Simulations of Oblique Pulsars}",
      journal = {\apjl},
         year = 2015,
        month = mar,
       volume = {801},
       number = {1},
          eid = {L19},
        pages = {L19},
          doi = {10.1088/2041-8205/801/1/L19},
archivePrefix = {arXiv},
       eprint = {1412.0673},
 primaryClass = {astro-ph.HE},
       adsurl = {https://ui.adsabs.harvard.edu/abs/2015ApJ...801L..19P}
}

@ARTICLE{philippov2015b,
       author = {{Philippov}, Alexander A. and {Cerutti}, Beno{\^\i}t and {Tchekhovskoy}, Alexander and {Spitkovsky}, Anatoly},
        title = "{Ab Initio Pulsar Magnetosphere: The Role of General Relativity}",
      journal = {\apjl},
         year = 2015,
        month = dec,
       volume = {815},
       number = {2},
          eid = {L19},
        pages = {L19},
          doi = {10.1088/2041-8205/815/2/L19},
archivePrefix = {arXiv},
       eprint = {1510.01734},
 primaryClass = {astro-ph.HE},
       adsurl = {https://ui.adsabs.harvard.edu/abs/2015ApJ...815L..19P}
}

@ARTICLE{meszaros2002,
       author = {{M{\'e}sz{\'a}ros}, P.},
        title = "{Theories of Gamma-Ray Bursts}",
      journal = {\araa},
         year = 2002,
        month = jan,
       volume = {40},
        pages = {137-169},
          doi = {10.1146/annurev.astro.40.060401.093821},
archivePrefix = {arXiv},
       eprint = {astro-ph/0111170},
 primaryClass = {astro-ph},
       adsurl = {https://ui.adsabs.harvard.edu/abs/2002ARA&A..40..137M}
}

@ARTICLE{arons1979,
       author = {{Arons}, J.},
        title = "{Some problems of pulsar physics or I'm madly in love with electricity}",
      journal = {\ssr},
         year = 1979,
        month = dec,
       volume = {24},
       number = {4},
        pages = {437-510},
          doi = {10.1007/BF00172212},
       adsurl = {https://ui.adsabs.harvard.edu/abs/1979SSRv...24..437A}
}

@ARTICLE{hirotani2021,
       author = {{Hirotani}, Kouichi and {Krasnopolsky}, Ruben and {Shang}, Hsien and {Nishikawa}, Ken-ichi and {Watson}, Michael},
        title = "{Two-dimensional Particle-in-cell Simulations of Axisymmetric Black Hole Magnetospheres}",
      journal = {\apj},
         year = 2021,
        month = feb,
       volume = {908},
       number = {1},
          eid = {88},
        pages = {88},
          doi = {10.3847/1538-4357/abd3a6},
archivePrefix = {arXiv},
       eprint = {2012.07229},
 primaryClass = {astro-ph.HE},
       adsurl = {https://ui.adsabs.harvard.edu/abs/2021ApJ...908...88H}
}

@ARTICLE{petri2009,
       author = {{P{\'e}tri}, J.},
        title = "{Non-linear evolution of the diocotron instability in a pulsar electrosphere: two-dimensional particle-in-cell simulations}",
      journal = {\aap},
         year = 2009,
        month = aug,
       volume = {503},
       number = {1},
        pages = {1-12},
          doi = {10.1051/0004-6361/200911778},
archivePrefix = {arXiv},
       eprint = {0905.1076},
 primaryClass = {astro-ph.HE},
       adsurl = {https://ui.adsabs.harvard.edu/abs/2009A&A...503....1P}
}

@ARTICLE{maero2024,
       author = {{Maero}, G. and {Hunter}, E.~D. and {Murtagh}, D.~J. and {Stenson}, E.~V.},
        title = "{Fundamental physics and other applications using nonneutral plasma}",
      journal = {Advances in Physics X},
         year = 2024,
        month = dec,
       volume = {9},
       number = {1},
          eid = {2367438},
        pages = {2367438},
          doi = {10.1080/23746149.2024.2367438},
       adsurl = {https://ui.adsabs.harvard.edu/abs/2024AdPhX...967438M}
}

@ARTICLE{chen2023,
       author = {{Chen}, Hui and {Fiuza}, Frederico},
        title = "{Perspectives on relativistic electron-positron pair plasma experiments of astrophysical relevance using high-power lasers}",
      journal = {Physics of Plasmas},
         year = 2023,
        month = feb,
       volume = {30},
       number = {2},
          eid = {020601},
        pages = {020601},
          doi = {10.1063/5.0134819},
       adsurl = {https://ui.adsabs.harvard.edu/abs/2023PhPl...30b0601C}
}

@ARTICLE{arrowsmith2024,
       author = {{Arrowsmith}, C.~D. and {Simon}, P. and {Bilbao}, P.~J. and {Bott}, A.~F.~A. and {Burger}, S. and {Chen}, H. and {Cruz}, F.~D. and {Davenne}, T. and {Efthymiopoulos}, I. and {Froula}, D.~H. and {Goillot}, A. and {Gudmundsson}, J.~T. and {Haberberger}, D. and {Halliday}, J.~W.~D. and {Hodge}, T. and {Huffman}, B.~T. and {Iaquinta}, S. and {Miniati}, F. and {Reville}, B. and {Sarkar}, S. and {Schekochihin}, A.~A. and {Silva}, L.~O. and {Simpson}, R. and {Stergiou}, V. and {Trines}, R.~M.~G.~M. and {Vieu}, T. and {Charitonidis}, N. and {Bingham}, R. and {Gregori}, G.},
        title = "{Laboratory realization of relativistic pair-plasma beams}",
      journal = {Nature Communications},
         year = 2024,
        month = jun,
       volume = {15},
          eid = {5029},
        pages = {5029},
          doi = {10.1038/s41467-024-49346-2},
archivePrefix = {arXiv},
       eprint = {2312.05244},
 primaryClass = {physics.plasm-ph},
       adsurl = {https://ui.adsabs.harvard.edu/abs/2024NatCo..15.5029A}
}

@ARTICLE{GJ69,
       author = {{Goldreich}, Peter and {Julian}, William H.},
        title = "{Pulsar Electrodynamics}",
      journal = {\apj},
         year = 1969,
        month = aug,
       volume = {157},
        pages = {869},
          doi = {10.1086/150119},
       adsurl = {https://ui.adsabs.harvard.edu/abs/1969ApJ...157..869G}
}

@ARTICLE{philippov-kramer22,
       author = {{Philippov}, A. and {Kramer}, M.},
        title = "{Pulsar Magnetospheres and Their Radiation}",
      journal = {\araa},
         year = 2022,
        month = aug,
       volume = {60},
        pages = {495-558},
          doi = {10.1146/annurev-astro-052920-112338},
       adsurl = {https://ui.adsabs.harvard.edu/abs/2022ARA&A..60..495P}
}

@article{timokhin10,
    author = {Timokhin, A. N.},
    title = {Time-dependent pair cascades in magnetospheres of neutron stars – I. Dynamics of the polar cap cascade with no particle supply from the neutron star surface},
    journal = {Monthly Notices of the Royal Astronomical Society},
    volume = {408},
    number = {4},
    pages = {2092-2114},
    year = {2010},
    month = {10},
    issn = {0035-8711},
    doi = {10.1111/j.1365-2966.2010.17286.x},
    url = {https://doi.org/10.1111/j.1365-2966.2010.17286.x},
    eprint = {https://academic.oup.com/mnras/article-pdf/408/4/2092/4220523/mnras0408-2092.pdf},
}

@ARTICLE{philippov-timokhin20,
       author = {{Philippov}, Alexander and {Timokhin}, Andrey and {Spitkovsky}, Anatoly},
        title = "{Origin of Pulsar Radio Emission}",
      journal = {\prl},
         year = 2020,
        month = jun,
       volume = {124},
       number = {24},
          eid = {245101},
        pages = {245101},
          doi = {10.1103/PhysRevLett.124.245101},
archivePrefix = {arXiv},
       eprint = {2001.02236},
 primaryClass = {astro-ph.HE},
       adsurl = {https://ui.adsabs.harvard.edu/abs/2020PhRvL.124x5101P}
}

@ARTICLE{sashaCh+24,
       author = {{Chernoglazov}, Alexander and {Philippov}, Alexander and {Timokhin}, Andrey},
        title = "{Coherence of Multidimensional Pair Production Discharges in Polar Caps of Pulsars}",
      journal = {\apjl},
         year = 2024,
        month = oct,
       volume = {974},
       number = {2},
          eid = {L32},
        pages = {L32},
          doi = {10.3847/2041-8213/ad7e24},
archivePrefix = {arXiv},
       eprint = {2409.15409},
 primaryClass = {astro-ph.HE},
       adsurl = {https://ui.adsabs.harvard.edu/abs/2024ApJ...974L..32C}
}

@ARTICLE{beloborodov-thompson07,
       author = {{Beloborodov}, Andrei M. and {Thompson}, Christopher},
        title = "{Corona of Magnetars}",
      journal = {\apj},
         year = 2007,
        month = mar,
       volume = {657},
       number = {2},
        pages = {967-993},
          doi = {10.1086/508917},
archivePrefix = {arXiv},
       eprint = {astro-ph/0602417},
 primaryClass = {astro-ph},
       adsurl = {https://ui.adsabs.harvard.edu/abs/2007ApJ...657..967B}
}

@ARTICLE{bransgrove+24,
       author = {{Bransgrove}, Ashley and {Levin}, Yuri and {Beloborodov}, Andrei M.},
        title = "{Giant Hall Waves Launched by Superconducting Phase Transition in Pulsars}",
      journal = {arXiv e-prints},
         year = 2024,
        month = aug,
          eid = {arXiv:2408.10888},
        pages = {arXiv:2408.10888},
          doi = {10.48550/arXiv.2408.10888},
archivePrefix = {arXiv},
       eprint = {2408.10888},
 primaryClass = {astro-ph.HE},
       adsurl = {https://ui.adsabs.harvard.edu/abs/2024arXiv240810888B}
}

@ARTICLE{potekhin+13,
       author = {{Potekhin}, A.~Y. and {Fantina}, A.~F. and {Chamel}, N. and {Pearson}, J.~M. and {Goriely}, S.},
        title = "{Analytical representations of unified equations of state for neutron-star matter}",
      journal = {\aap},
         year = 2013,
        month = dec,
       volume = {560},
          eid = {A48},
        pages = {A48},
          doi = {10.1051/0004-6361/201321697},
archivePrefix = {arXiv},
       eprint = {1310.0049},
 primaryClass = {astro-ph.SR},
       adsurl = {https://ui.adsabs.harvard.edu/abs/2013A&A...560A..48P}
}

@ARTICLE{pearson+11,
       author = {{Pearson}, J.~M. and {Goriely}, S. and {Chamel}, N.},
        title = "{Properties of the outer crust of neutron stars from Hartree-Fock-Bogoliubov mass models}",
      journal = {\prc},
         year = 2011,
        month = jun,
       volume = {83},
       number = {6},
          eid = {065810},
        pages = {065810},
          doi = {10.1103/PhysRevC.83.065810},
       adsurl = {https://ui.adsabs.harvard.edu/abs/2011PhRvC..83f5810P}
}

@ARTICLE{groselj2018,
       author = {{Gro{\v{s}}elj}, Daniel and {Mallet}, Alfred and {Loureiro}, Nuno F. and {Jenko}, Frank},
        title = "{Fully Kinetic Simulation of 3D Kinetic Alfv{\'e}n Turbulence}",
      journal = {\prl},
         year = 2018,
        month = mar,
       volume = {120},
       number = {10},
          eid = {105101},
        pages = {105101},
          doi = {10.1103/PhysRevLett.120.105101},
archivePrefix = {arXiv},
       eprint = {1710.03581},
 primaryClass = {physics.plasm-ph},
       adsurl = {https://ui.adsabs.harvard.edu/abs/2018PhRvL.120j5101G}
}

@ARTICLE{arzamasskiy2019,
       author = {{Arzamasskiy}, Lev and {Kunz}, Matthew W. and {Chandran}, Benjamin D.~G. and {Quataert}, Eliot},
        title = "{Hybrid-kinetic Simulations of Ion Heating in Alfv{\'e}nic Turbulence}",
      journal = {\apj},
         year = 2019,
        month = jul,
       volume = {879},
       number = {1},
          eid = {53},
        pages = {53},
          doi = {10.3847/1538-4357/ab20cc},
archivePrefix = {arXiv},
       eprint = {1901.11028},
 primaryClass = {astro-ph.HE},
       adsurl = {https://ui.adsabs.harvard.edu/abs/2019ApJ...879...53A}
}

@ARTICLE{cerri2018,
       author = {{Cerri}, S.~S. and {Kunz}, M.~W. and {Califano}, F.},
        title = "{Dual Phase-space Cascades in 3D Hybrid-Vlasov-Maxwell Turbulence}",
      journal = {\apjl},
         year = 2018,
        month = mar,
       volume = {856},
       number = {1},
          eid = {L13},
        pages = {L13},
          doi = {10.3847/2041-8213/aab557},
archivePrefix = {arXiv},
       eprint = {1802.06133},
 primaryClass = {physics.plasm-ph},
       adsurl = {https://ui.adsabs.harvard.edu/abs/2018ApJ...856L..13C}
}

@ARTICLE{servidio2015,
       author = {{Servidio}, S. and {Valentini}, F. and {Perrone}, D. and {Greco}, A. and {Califano}, F. and {Matthaeus}, W.~H. and {Veltri}, P.},
        title = "{A kinetic model of plasma turbulence}",
      journal = {Journal of Plasma Physics},
         year = 2015,
        month = jan,
       volume = {81},
       number = {1},
          eid = {325810107},
        pages = {325810107},
          doi = {10.1017/S0022377814000841},
       adsurl = {https://ui.adsabs.harvard.edu/abs/2015JPlPh..81a3207S}
}

@ARTICLE{leamon1998,
       author = {{Leamon}, Robert J. and {Smith}, Charles W. and {Ness}, Norman F. and {Matthaeus}, William H. and {Wong}, Hung K.},
        title = "{Observational constraints on the dynamics of the interplanetary magnetic field dissipation range}",
      journal = {\jgr},
         year = 1998,
        month = mar,
       volume = {103},
       number = {A3},
        pages = {4775-4788},
          doi = {10.1029/97JA03394},
       adsurl = {https://ui.adsabs.harvard.edu/abs/1998JGR...103.4775L}
}

@ARTICLE{salem2012,
       author = {{Salem}, C.~S. and {Howes}, G.~G. and {Sundkvist}, D. and {Bale}, S.~D. and {Chaston}, C.~C. and {Chen}, C.~H.~K. and {Mozer}, F.~S.},
        title = "{Identification of Kinetic Alfv{\'e}n Wave Turbulence in the Solar Wind}",
      journal = {\apjl},
         year = 2012,
        month = jan,
       volume = {745},
       number = {1},
          eid = {L9},
        pages = {L9},
          doi = {10.1088/2041-8205/745/1/L9},
       adsurl = {https://ui.adsabs.harvard.edu/abs/2012ApJ...745L...9S}
}

@ARTICLE{sahraoui2009,
       author = {{Sahraoui}, F. and {Goldstein}, M.~L. and {Robert}, P. and {Khotyaintsev}, Yu. V.},
        title = "{Evidence of a Cascade and Dissipation of Solar-Wind Turbulence at the Electron Gyroscale}",
      journal = {\prl},
         year = 2009,
        month = jun,
       volume = {102},
       number = {23},
          eid = {231102},
        pages = {231102},
          doi = {10.1103/PhysRevLett.102.231102},
       adsurl = {https://ui.adsabs.harvard.edu/abs/2009PhRvL.102w1102S}
}

@ARTICLE{biskamp1999,
       author = {{Biskamp}, D. and {Schwarz}, E. and {Zeiler}, A. and {Celani}, A. and {Drake}, J.~F.},
        title = "{Electron magnetohydrodynamic turbulence}",
      journal = {Physics of Plasmas},
         year = 1999,
        month = mar,
       volume = {6},
       number = {3},
        pages = {751-758},
          doi = {10.1063/1.873312},
       adsurl = {https://ui.adsabs.harvard.edu/abs/1999PhPl....6..751B}
}

@ARTICLE{cho2009,
       author = {{Cho}, Jungyeon and {Lazarian}, A.},
        title = "{Simulations of Electron Magnetohydrodynamic Turbulence}",
      journal = {\apj},
         year = 2009,
        month = aug,
       volume = {701},
       number = {1},
        pages = {236-252},
          doi = {10.1088/0004-637X/701/1/236},
archivePrefix = {arXiv},
       eprint = {0904.0661},
 primaryClass = {astro-ph.EP},
       adsurl = {https://ui.adsabs.harvard.edu/abs/2009ApJ...701..236C}
}

@ARTICLE{zhou2023,
       author = {{Zhou}, Muni and {Liu}, Zhuo and {Loureiro}, Nuno F.},
        title = "{Spectrum of kinetic-Alfv{\'e}n-wave turbulence: intermittency or tearing mediation?}",
      journal = {\mnras},
         year = 2023,
        month = oct,
       volume = {524},
       number = {4},
        pages = {5468-5476},
          doi = {10.1093/mnras/stad2231},
archivePrefix = {arXiv},
       eprint = {2208.02441},
 primaryClass = {astro-ph.HE},
       adsurl = {https://ui.adsabs.harvard.edu/abs/2023MNRAS.524.5468Z}
}

@ARTICLE{urpin2011,
       author = {{Urpin}, V.},
        title = "{Magnetohydrodynamic waves in the pulsar magnetosphere}",
      journal = {\aap},
         year = 2011,
        month = nov,
       volume = {535},
          eid = {L5},
        pages = {L5},
          doi = {10.1051/0004-6361/201118047},
archivePrefix = {arXiv},
       eprint = {1110.6302},
 primaryClass = {astro-ph.SR},
       adsurl = {https://ui.adsabs.harvard.edu/abs/2011A&A...535L...5U}
}

@ARTICLE{gedalin2001,
       author = {{Gedalin}, M. and {Gruman}, E. and {Melrose}, D.~B.},
        title = "{Low-frequency waves in asymmetric magnetized relativistic pair plasma}",
      journal = {\mnras},
         year = 2001,
        month = aug,
       volume = {325},
       number = {2},
        pages = {715-725},
          doi = {10.1046/j.1365-8711.2001.04483.x},
       adsurl = {https://ui.adsabs.harvard.edu/abs/2001MNRAS.325..715G}
}

@ARTICLE{varma2011,
       author = {{Agarwal}, P. and {Varma}, P. and {Tiwari}, M.~S.},
        title = "{Study of inertial kinetic Alfven waves around cusp region}",
      journal = {\planss},
         year = 2011,
        month = mar,
       volume = {59},
       number = {4},
        pages = {306-311},
          doi = {10.1016/j.pss.2010.11.006},
       adsurl = {https://ui.adsabs.harvard.edu/abs/2011P&SS...59..306A}
}

@ARTICLE{shukla2009,
       author = {{Shukla}, N. and {Varma}, P. and {Tivari}, M. S.},
        title = "{Study on kinetic Alfven wave in inertial regime}",
      journal = {Indian Journal of Pure \& Applied Physics},
         year = 2009,
        month = may,
       volume = {47},
       number = {},
          eid = {},
        pages = {350},
          doi = {},
archivePrefix = {},
       eprint = {},
 primaryClass = {},
       adsurl = {}
}

@ARTICLE{mallet2015,
       author = {{Mallet}, A. and {Schekochihin}, A.~A. and {Chandran}, B.~D.~G.},
        title = "{Refined critical balance in strong Alfvenic turbulence.}",
      journal = {\mnras},
         year = 2015,
        month = apr,
       volume = {449},
        pages = {L77-L81},
          doi = {10.1093/mnrasl/slv021},
archivePrefix = {arXiv},
       eprint = {1406.5658},
 primaryClass = {astro-ph.SR},
       adsurl = {https://ui.adsabs.harvard.edu/abs/2015MNRAS.449L..77M}
}

@ARTICLE{vega2024,
       author = {{Vega}, Cristian and {Boldyrev}, Stanislav and {Roytershteyn}, Vadim},
        title = "{Relativistic Alfv{\'e}n Turbulence at Kinetic Scales}",
      journal = {\apj},
         year = 2024,
        month = apr,
       volume = {965},
       number = {1},
          eid = {27},
        pages = {27},
          doi = {10.3847/1538-4357/ad2e02},
archivePrefix = {arXiv},
       eprint = {2402.16218},
 primaryClass = {physics.plasm-ph},
       adsurl = {https://ui.adsabs.harvard.edu/abs/2024ApJ...965...27V}
}

@ARTICLE{told2015,
       author = {{Told}, D. and {Jenko}, F. and {TenBarge}, J.~M. and {Howes}, G.~G. and {Hammett}, G.~W.},
        title = "{Multiscale Nature of the Dissipation Range in Gyrokinetic Simulations of Alfv{\'e}nic Turbulence}",
      journal = {\prl},
         year = 2015,
        month = jul,
       volume = {115},
       number = {2},
          eid = {025003},
        pages = {025003},
          doi = {10.1103/PhysRevLett.115.025003},
archivePrefix = {arXiv},
       eprint = {1505.02204},
 primaryClass = {physics.plasm-ph},
       adsurl = {https://ui.adsabs.harvard.edu/abs/2015PhRvL.115b5003T}
}

@ARTICLE{arons1986,
       author = {{Arons}, J. and {Barnard}, J.~J.},
        title = "{Wave Propagation in Pulsar Magnetospheres: Dispersion Relations and Normal Modes of Plasmas in Superstrong Magnetic Fields}",
      journal = {\apj},
         year = 1986,
        month = mar,
       volume = {302},
        pages = {120},
          doi = {10.1086/163978},
       adsurl = {https://ui.adsabs.harvard.edu/abs/1986ApJ...302..120A}
}

@ARTICLE{gedalin1998,
       author = {{Gedalin}, M. and {Melrose}, D.~B. and {Gruman}, E.},
        title = "{Long waves in a relativistic pair plasma in a strong magnetic field}",
      journal = {\pre},
         year = 1998,
        month = mar,
       volume = {57},
       number = {3},
        pages = {3399-3410},
          doi = {10.1103/PhysRevE.57.3399},
       adsurl = {https://ui.adsabs.harvard.edu/abs/1998PhRvE..57.3399G}
}

@ARTICLE{godfrey1975,
       author = {{Godfrey}, B.~B. and {Newberger}, B.~S. and {Taggart}, K.~A.},
        title = "{A relativistic plasma dispersion function}",
      journal = {IEEE Transactions on Plasma Science},
         year = 1975,
        month = jun,
       volume = {3},
        pages = {60-67},
          doi = {10.1109/TPS.1975.4316876},
       adsurl = {https://ui.adsabs.harvard.edu/abs/1975ITPS....3...60G}
}

@ARTICLE{chernoglazov2021,
       author = {{Chernoglazov}, Alexander and {Ripperda}, Bart and {Philippov}, Alexander},
        title = "{Dynamic Alignment and Plasmoid Formation in Relativistic Magnetohydrodynamic Turbulence}",
      journal = {\apjl},
         year = 2021,
        month = dec,
       volume = {923},
       number = {1},
          eid = {L13},
        pages = {L13},
          doi = {10.3847/2041-8213/ac3afa},
archivePrefix = {arXiv},
       eprint = {2111.08188},
 primaryClass = {astro-ph.HE},
       adsurl = {https://ui.adsabs.harvard.edu/abs/2021ApJ...923L..13C}
}

@ARTICLE{vega2022b,
       author = {{Vega}, Cristian and {Boldyrev}, Stanislav and {Roytershteyn}, Vadim},
        title = "{Spectra of Magnetic Turbulence in a Relativistic Plasma}",
      journal = {\apjl},
         year = 2022,
        month = may,
       volume = {931},
       number = {1},
          eid = {L10},
        pages = {L10},
          doi = {10.3847/2041-8213/ac6cde},
archivePrefix = {arXiv},
       eprint = {2204.04530},
 primaryClass = {physics.plasm-ph},
       adsurl = {https://ui.adsabs.harvard.edu/abs/2022ApJ...931L..10V}
}

@article{kasper2021,
  title = {Parker Solar Probe Enters the Magnetically Dominated Solar Corona},
  author = {Kasper, J. C. and Klein, K. G. and Lichko, E. and Huang, Jia and Chen, C. H. K. and Badman, S. T. and Bonnell, J. and Whittlesey, P. L. and Livi, R. and Larson, D. and Pulupa, M. and Rahmati, A. and Stansby, D. and Korreck, K. E. and Stevens, M. and Case, A. W. and Bale, S. D. and Maksimovic, M. and Moncuquet, M. and Goetz, K. and Halekas, J. S. and Malaspina, D. and Raouafi, Nour E. and Szabo, A. and MacDowall, R. and Velli, Marco and Dudok de Wit, Thierry and Zank, G. P.},
  journal = {Phys. Rev. Lett.},
  volume = {127},
  issue = {25},
  pages = {255101},
  numpages = {8},
  year = {2021},
  month = {Dec},
  publisher = {American Physical Society},
  doi = {10.1103/PhysRevLett.127.255101},
  url = {https://link.aps.org/doi/10.1103/PhysRevLett.127.255101}
}

@ARTICLE{milanese2020,
       author = {{Milanese}, Lucio M. and {Loureiro}, Nuno F. and {Daschner}, Maximilian and {Boldyrev}, Stanislav},
        title = "{Dynamic Phase Alignment in Inertial Alfv{\'e}n Turbulence}",
      journal = {\prl},
         year = 2020,
        month = dec,
       volume = {125},
       number = {26},
          eid = {265101},
        pages = {265101},
          doi = {10.1103/PhysRevLett.125.265101},
archivePrefix = {arXiv},
       eprint = {2010.00415},
 primaryClass = {physics.plasm-ph},
       adsurl = {https://ui.adsabs.harvard.edu/abs/2020PhRvL.125z5101M}
}

@ARTICLE{boldyrev2006,
       author = {{Boldyrev}, Stanislav},
        title = "{Spectrum of Magnetohydrodynamic Turbulence}",
      journal = {\prl},
         year = 2006,
        month = mar,
       volume = {96},
       number = {11},
          eid = {115002},
        pages = {115002},
          doi = {10.1103/PhysRevLett.96.115002},
archivePrefix = {arXiv},
       eprint = {astro-ph/0511290},
 primaryClass = {astro-ph},
       adsurl = {https://ui.adsabs.harvard.edu/abs/2006PhRvL..96k5002B}
}

@ARTICLE{mason2006,
       author = {{Mason}, Joanne and {Cattaneo}, Fausto and {Boldyrev}, Stanislav},
        title = "{Dynamic Alignment in Driven Magnetohydrodynamic Turbulence}",
      journal = {\prl},
         year = 2006,
        month = dec,
       volume = {97},
       number = {25},
          eid = {255002},
        pages = {255002},
          doi = {10.1103/PhysRevLett.97.255002},
archivePrefix = {arXiv},
       eprint = {astro-ph/0602382},
 primaryClass = {astro-ph},
       adsurl = {https://ui.adsabs.harvard.edu/abs/2006PhRvL..97y5002M}
}

@ARTICLE{boldyrev2021,
       author = {{Boldyrev}, Stanislav and {Loureiro}, Nuno F. and {Roytershteyn}, Vadim},
        title = "{Plasma dynamics in low-electron-beta environments}",
      journal = {Frontiers in Astronomy and Space Sciences},
         year = 2021,
        month = may,
       volume = {8},
          eid = {15},
        pages = {15},
          doi = {10.3389/fspas.2021.621040},
       adsurl = {https://ui.adsabs.harvard.edu/abs/2021FrASS...8...15B}
}

@ARTICLE{loureiro2018,
       author = {{Loureiro}, Nuno F. and {Boldyrev}, Stanislav},
        title = "{Turbulence in Magnetized Pair Plasmas}",
      journal = {\apjl},
         year = 2018,
        month = oct,
       volume = {866},
       number = {1},
          eid = {L14},
        pages = {L14},
          doi = {10.3847/2041-8213/aae483},
archivePrefix = {arXiv},
       eprint = {1805.09224},
 primaryClass = {physics.plasm-ph},
       adsurl = {https://ui.adsabs.harvard.edu/abs/2018ApJ...866L..14L}
}

@ARTICLE{tenbarge2012,
   author = {{TenBarge}, J.~M. and {Howes}, G.~G.},
    title = "{Evidence of critical balance in kinetic Alfv{\'e}n wave turbulence simulations a)}",
  journal = {Physics of Plasmas},
archivePrefix = "arXiv",
   eprint = {1201.0056},
 primaryClass = "physics.plasm-ph",
     year = 2012,
    month = may,
   volume = 19,
   number = 5,
      eid = {055901},
    pages = {055901},
      doi = {10.1063/1.3693974},
   adsurl = {http://adsabs.harvard.edu/abs/2012PhPl...19e5901T}
}

@article{chen_boldyrev2017,
       author = {{Chen}, C.~H.~K. and {Boldyrev}, S.},
        title = "{Nature of Kinetic Scale Turbulence in the Earth's Magnetosheath}",
      journal = {\apj},
         year = "2017",
        month = "Jun",
       volume = {842},
       number = {2},
          eid = {122},
        pages = {122},
          doi = {10.3847/1538-4357/aa74e0},
archivePrefix = {arXiv},
       eprint = {1705.08558},
 primaryClass = {physics.space-ph},
       adsurl = {https://ui.adsabs.harvard.edu/abs/2017ApJ...842..122C}
}

@ARTICLE{walker2018,
   author = {{Walker}, J. and {Boldyrev}, S. and {Loureiro}, N.~F.},
    title = "{Influence of tearing instability on magnetohydrodynamic turbulence}",
  journal = {Physical Review E},
archivePrefix = "arXiv",
   eprint = {1804.02754},
 primaryClass = "physics.plasm-ph",
     year = 2018,
    month = sep,
   volume = 98,
   number = 3,
      eid = {033209},
    pages = {033209},
      doi = {10.1103/PhysRevE.98.033209},
   adsurl = {http://adsabs.harvard.edu/abs/2018PhRvE..98c3209W}
}

@ARTICLE{chen2016,
   author = {{Chen}, C.~H.~K.},
    title = "{Recent progress in astrophysical plasma turbulence from solar wind observations}",
  journal = {Journal of Plasma Physics},
archivePrefix = "arXiv",
   eprint = {1611.03386},
 primaryClass = "physics.plasm-ph",
     year = 2016,
    month = dec,
   volume = 82,
   number = 6,
      eid = {535820602},
    pages = {535820602},
      doi = {10.1017/S0022377816001124},
   adsurl = {http://adsabs.harvard.edu/abs/2016JPlPh..82f5302C}
}

@ARTICLE{wan2015,
       author = {{Wan}, M. and {Matthaeus}, W.~H. and {Roytershteyn}, V. and {Karimabadi}, H. and {Parashar}, T. and {Wu}, P. and {Shay}, M.},
        title = "{Intermittent Dissipation and Heating in 3D Kinetic Plasma Turbulence}",
      journal = {\prl},
         year = 2015,
        month = may,
       volume = {114},
       number = {17},
          eid = {175002},
        pages = {175002},
          doi = {10.1103/PhysRevLett.114.175002},
       adsurl = {https://ui.adsabs.harvard.edu/abs/2015PhRvL.114q5002W}
}

@ARTICLE{boldyrev12b,
   author = {{Boldyrev}, S. and {Perez}, J.~C.},
    title = "{Spectrum of Kinetic-Alfv{\'e}n Turbulence}",
  journal = {The Astrophysical Journal},
     year = 2012,
    month = oct,
   volume = 758,
      eid = {L44},
    pages = {L44},
      doi = {10.1088/2041-8205/758/2/L44},
   adsurl = {http://adsabs.harvard.edu/abs/2012ApJ...758L..44B}
}

@BOOK{tobias2013,
   author = {{Tobias}, S.~M. and {Cattaneo}, F. and {Boldyrev}, S.},
    title = "{MHD Dynamos and Turbulence, {\em in} Ten Chapters in Turbulence, ed. P.A. Davidson, Y. Kaneda, and K.R. Sreenivasan : Cambridge University Press, p. 351-404. }",	
booktitle = {Ten Chapters in Turbulence, ed. P.A. Davidson, Y. Kaneda, and K.R. Sreenivasan : Cambridge University Press, p. 351-404, (2013).},    
     year = 2013,
	archivePrefix = "arXiv",
   eprint = {1103.3138},
    month = {},
    pages = {}, 
   adsurl = {http://adsabs.harvard.edu/abs/2011arXiv1103.3138T}
}

@BOOK{biskamp2003,
   author = {{Biskamp}, D.},
    title = "{Magnetohydrodynamic Turbulence,~Cambridge, UK: Cambridge University Press.}",
    publisher={Cambridge University Press},
     year = 2003,
    month = jul,
    pages = {310},
   adsurl = {http://adsabs.harvard.edu/abs/2003matu.book.....B}
}

@ARTICLE{mason2012,
   author = {{Mason}, J. and {Perez}, J.~C. and {Boldyrev}, S. and {Cattaneo}, F.
	},
    title = "{Numerical simulations of strong incompressible magnetohydrodynamic turbulence}",
  journal = {Physics of Plasmas},
archivePrefix = "arXiv",
   eprint = {1202.3474},
 primaryClass = "physics.plasm-ph",
     year = 2012,
    month = may,
   volume = 19,
   number = 5,
    pages = {055902-055902},
      doi = {10.1063/1.3694123},
   adsurl = {http://adsabs.harvard.edu/abs/2012PhPl...19e5902M}
}

@ARTICLE{boldyrev2009,
   author = {{Boldyrev}, S. and {Mason}, J. and {Cattaneo}, F.},
    title = "{Dynamic Alignment and Exact Scaling Laws in Magnetohydrodynamic Turbulence}",
  journal = {The Astrophysical Journal Letters},
     year = 2009,
    month = jul,
   volume = 699,
    pages = {L39-L42},
      doi = {10.1088/0004-637X/699/1/L39},
   adsurl = {http://adsabs.harvard.edu/abs/2009ApJ...699L..39B}
}

@ARTICLE{perez_etal2012,
   author = {{Perez}, J.~C. and {Mason}, J. and {Boldyrev}, S. and {Cattaneo}, F.
	},
    title = "{On the Energy Spectrum of Strong Magnetohydrodynamic Turbulence}",
  journal = {Physical Review X},
archivePrefix = "arXiv",
   eprint = {1209.2011},
 primaryClass = "astro-ph.SR",
     year = 2012,
    month = oct,
   volume = 2,
   number = 4,
      eid = {041005},
    pages = {041005},
      doi = {10.1103/PhysRevX.2.041005},
   adsurl = {http://adsabs.harvard.edu/abs/2012PhRvX...2d1005P}
}

@ARTICLE{boldyrev_p12,
   author = {{Boldyrev}, S. and {Perez}, J.~C.},
    title = "{Spectrum of Kinetic-Alfv{\'e}n Turbulence}",
  journal = {The Astrophysical Journal Letters},
archivePrefix = "arXiv",
   eprint = {1204.5809},
 primaryClass = "astro-ph.SR",
     year = 2012,
    month = oct,
   volume = 758,
      eid = {L44},
    pages = {L44},
      doi = {10.1088/2041-8205/758/2/L44},
   adsurl = {http://adsabs.harvard.edu/abs/2012ApJ...758L..44B}
}

@ARTICLE{goldreich_toward_1995,
   author = {{Goldreich}, P. and {Sridhar}, S.},
    title = "{Toward a theory of interstellar turbulence. 2: Strong alfvenic turbulence}",
  journal = {The Astrophysical Journal},
     year = 1995,
    month = jan,
   volume = 438,
    pages = {763-775},
      doi = {10.1086/175121},
   adsurl = {http://adsabs.harvard.edu/abs/1995ApJ...438..763G}
}

@ARTICLE{bale_measurement_2005,
   author = {{Bale}, S.~D. and {Kellogg}, P.~J. and {Mozer}, F.~S. and {Horbury}, T.~S. and 
{Reme}, H.},
    title = "{Measurement of the Electric Fluctuation Spectrum of Magnetohydrodynamic Turbulence}",
  journal = {Physical Review Letters},
   eprint = {physics/0503103},
     year = 2005,
    month = jun,
   volume = 94,
   number = 21,
      eid = {215002},
    pages = {215002},
      doi = {10.1103/PhysRevLett.94.215002},
   adsurl = {http://adsabs.harvard.edu/abs/2005PhRvL..94u5002B}
}

@ARTICLE{chandran_intermittency_2015,
   author = {{Chandran}, B.~D.~G. and {Schekochihin}, A.~A. and {Mallet}, A.
},
    title = "{Intermittency and Alignment in Strong RMHD Turbulence}",
  journal = {The Astrophysical Journal},
archivePrefix = "arXiv",
   eprint = {1403.6354},
 primaryClass = "astro-ph.SR",
     year = 2015,
    month = jul,
   volume = 807,
      eid = {39},
    pages = {39},
      doi = {10.1088/0004-637X/807/1/39},
   adsurl = {http://adsabs.harvard.edu/abs/2015ApJ...807...39C}
}
\bibliographystyle{aasjournal}



\end{document}